\documentclass[aps, prd, twocolumn, showpacs, superscriptaddress, groupedaddress]{revtex4-1} 
\usepackage{graphicx}	
\usepackage{amssymb}
\usepackage{dcolumn}
\usepackage{mathtools}
\usepackage{amsmath}
\usepackage{xcolor}
\usepackage{color}
\usepackage{subfigure, rotating, bm, array}
\usepackage[pagebackref=false, colorlinks=true]{hyperref}
\hypersetup{linkcolor=blue, citecolor=blue,
urlcolor=blue} 

\begin{document}

\title{Rotational energy extraction from the Kerr black hole's mimickers}
\author{Vishva Patel}
\email{vishwapatel2550@gmail.com}
\affiliation{International Center for Cosmology, Charusat University, Anand, GUJ 388421, India}
\author{Kauntey Acharya}
\email{kaunteyacharya2000@gmail.com}
\affiliation{International Center for Cosmology, Charusat University, Anand, GUJ 388421, India}
\author{Parth Bambhaniya}
\email{grcollapse@gmail.com}
\affiliation{International Center for Cosmology, Charusat University, Anand, GUJ 388421, India}
\author{Pankaj S. Joshi}
\email{psjcosmos@gmail.com}
\affiliation{International Centre for Space and Cosmology, Ahmedabad University, Ahmedabad, GUJ 380009, India}
\affiliation{International Center for Cosmology, Charusat University, Anand, GUJ 388421, India}
\date{\today}

\begin{abstract}
In this paper, the Penrose process is being used to extract rotational energy from regular black holes. Initially, we consider the rotating Simpson-Visser regular spacetime which describes the class of geometries of the Kerr black hole's mimickers. The Penrose process is then studied through conformally transformed rotating singular and regular black hole solutions. These both Simpson-Visser and conformally transformed geometries depend on mass, spin, and an additional regularisation parameter $l$. In both cases, we investigate how the spin and regularisation parameter $l$ affects the configuration of an ergoregion and event horizons. Surprisingly, we find that the energy extraction efficiency from the event horizon surface is not dependent on the regularisation parameter $l$ in the Simpson-Visser regular spacetimes and hence it does not vary from the Kerr black hole case. While, in the conformally transformed singular and regular black holes, we obtain the efficiency rate of extracted energies are extremely high compared to the Kerr black hole scenario. This distinct signature of the conformally transformed singular and regular black holes would be useful to distinguish them from the Kerr black hole in observation.

\end{abstract}
\maketitle

\section{Introduction}

The way mankind has witnessed a series of scientific breakthroughs in astrophysics, including the detection of gravitational waves from the merger of two black holes, the shadow images of M87* and Milky way galactic center Sgr- $A^{*}$ \cite{EventHorizonTelescope:2022xnr,EventHorizonTelescope:2022tzy,EventHorizonTelescope:2022ago,EventHorizonTelescope:2022vjs,EventHorizonTelescope:2022wok,EventHorizonTelescope:2022urf,EventHorizonTelescope:2022xqj,EventHorizonTelescope:2022gsd}, it has drawn attention of not just scientific community but also general public. These observations by the EHT group have opened up the way for gravitational theories to be verified in strong gravity regimes. The EHT group's findings have been used to constrain and study various aspects of gravity theories, extending from general relativity to its alternatives.

The stars having more mass than about 10 solar masses collapse under their own gravity at the end of their lives, since they cannot balance their gravitational pull by any known quantum pressures. Thus according to Einstein's general theory of relativity, the continuous gravitational collapse of a massive star culminates in a spacetime singularity that contains infinite matter density and gravitational field. At this point, all physical quantities diverge and become arbitrarily large. This ultra-dense region can be anticipated by modeling various astrophysical compact objects, such as black holes, naked singularities, worm holes and other specific types of these objects. These compact objects, together with gravitational waves and shadows, are considered one of the most efficient sources of energy in the Universe. As a result, they are assumed to be responsible for a gigantic electromagnetic environment in their near vicinity, as well as high-energy jet emission outbursts destroying nearby stars and galaxies.

In various literature, researchers have explored the different observational properties of different possible compact objects (Black holes, naked singularities, wormholes) for e.g. the shadow properties, \cite{Gralla:2019xty,Abdikamalov:2019ztb,Dey:2013yga,Dey:2020haf,Ohgami:2015nra,Sakai:2014pga,Saurabh:2022jjv,Solanki:2021mkt,Bambhaniya:2021ybs,Bambi:2015kza,Shaikh:2018kfv,Jusufi:2020zln}, gravitational lensing \cite{Shaikh:2019jfr,Shaikh:2018oul,Paul:2020ufc,Virbhadra:2007kw,Gyulchev:2008ff,Kala:2020prt,Sahu:2012er}, accretion disk properties \cite{Liu:2020vkh,Joshi:2013dva,Bambhaniya:2021ugr,Rahaman:2021kge,Harko:2009gc,Tahelyani:2022uxw,Guo:2020tgv,Chowdhury:2011aa} and orbital precession \cite{Martinez,Madan:2022spd, Eva1, Eva2, tsirulev, Joshi:2019rdo, Bambhaniya:2019pbr, Dey:2019fpv, Bam2020}. The similar observable properties were studied in singularity-free compact objects (regular black holes and worm holes) \cite{Guo:2022yjc,Kala:2022uog,Ahmed:2020ifa,Jafarzade:2021umv,Qin:2020xzu,Mondal:2020pop,Ghosh:2020ece,Kumar:2020ltt,Li:2013jra,Narzilloev:2020qtd,Benavides-Gallego:2018htf,Fathi:2020agx}. The major problem of powering active galactic nuclei, X-ray binaries, and quasars are the most important issues today in high-energy astrophysics. Several mechanisms have been proposed by various authors in different scenarios to explain these high energetic phenomena \cite{aa,ab}. 

In 1969, Roger Penrose presented an innovative and novel approach to extract energy from a rotating black hole. The Penrose process is dependent on the existence of an ergosphere, which is described as the region between an event horizon and the static surface limit. Since Penrose \cite{Penrose:1971uk} did not mention astrophysical implications of the Penrose process, Wheeler \cite{ac} and others stated that the process could offer a reasonable solution for high-energy jets coming from active galactic nuclei. This mechanism is known for a star-like object to approach a supermassive compact object and afterward break up into fragments due to immense tidal forces. Some fragments may have negative energy orbits, while others escape at extremely high velocity, generating a jet. As a consequence, the Penrose process has been reintroduced as a mechanism for high-energy sources. After a thorough investigation, multiple approaches (super Penrose process, collisional process, BSW (Banados-Silk-West) effect) for various scenarios were proposed \cite{Ghosh:2013ona,Liu:2012qe,Nozawa:2005eu,Wagh:1989zqa,Zaslavskii:2014eda1,Zaslavskii:2012ax1}. In \cite{Zaslavskii:2022nbm,Zaslavskii:2020pvu,Gupta:2021vww,Zaslavskii:2019pdc,Zaslavskii:2020fmz}, authors have studied the Super-Penrose process with charged particles near-naked singularity, white holes and wormholes. Researchers have also studied energy extraction in different scenarios like extremal rotating electrovacuum black holes using particle collision \cite{Hejda:2021cbk,Hejda:2019luo,Zaslavskii:2020kpv,Zaslavskii:2019pdca,Pavlov:2018kgs,Zaslavskii:2016unn,Zaslavskii:2015vaa}, and the Penrose process in axially symmetric magnetized black holes \cite{Shaymatov:2022eyz}.

As we all know, one of the fundamental problems in physics is the emergence of a curvature singularity within the general theory of relativity. Under certain physically reasonable conditions, they reflect the regions where geodesics abruptly terminate, and their occurrence is usually interpreted as indicating the theory's breakdown. It is generally speculated that quantum gravitational effects will prevent singularities from forming as an end state of gravitational collapse. However, a reasonable interpretation of such small-scale effects remains unclear. There have been various attempts to overcome the occurrence of a singularity \cite{a,Lawrence:1995ct,Easson:2003ia,Husain:2004yz,DeRoo:2010bqa,Cheng-Zhou:2010msa,HoffervLoewenfeld:2010vej,Singh:2012zzj,Corichi:2012xr,Lee:2012ju,Kreienbuehl:2013toa,Blanchette:2020kkk,Mosani:2021czj}. Motivated by this, in this paper, we consider the singularity-free solutions to study the Penrose process.

Initially, we consider a family of spacetime geometries proposed by Simpson and Visser \cite{Shaikh:2021yux,Mazza:2021rgq} that contains a class of solutions (one way wormhole, two-way wormhole and regular black hole) depending on the regularisation parameter $l$. Then we consider the conformally transformed solutions presented in Ref. \cite{Bambi:2016wdn}, where singularity-free black hole solutions have been proposed within a broad class of conformally invariant gravity theories. They have proposed singular and regular black hole solutions. These conformally transformed black hole spacetimes are solution of Conformal vacuum Einstein Field Equations (CEFE). For a regular black hole solution at r = 0, the curvature invariants do not diverge, demonstrating that the proposed spacetimes are geodesically complete \cite{Zhang:2018qdk}. The action in a conformally invariant theory of gravity is invariant under both generalized and conformal coordinate transformations. 

As mentioned above the observational properties have been studied in various literature for different non-singular compact objects. Moreover, the phenomenology of energy extraction has been studied in many cases where the compact objects could have horizons and a central singularity. It is, therefore, worth to study the energy extraction from a non-singular compact objects. Therefore, in this paper, we have consider the conformal rotating singular and non-singular spacetimes along with regular Simpson-Visser metrics, which we will use to study the phenomenology of the Penrose process. 

This paper is assembled as follows. In Section (\ref{PP}), we obtain the general formalism of the Penrose process. In the Section (\ref{SVsection}), we discuss the energy extraction rate from Simpson- Visser spacetimes. In Section (\ref{regusingusection}), we discuss the singular and regular black hole solutions and compare the efficiency of extracted energies with the Kerr black hole case. Finally, in Section (\ref{conclusion}), we wrap up the study and discuss the results. Throughout the paper, we have considered geometrized units. Thus, the gravitational constant (G) and the speed of light (c) are set equal to one. The signature of the metric is considered as (-,+,+,+).

\section{The Penrose process for energy extraction from rotating black hole}
\label{PP}
The Penrose process, which allows us to extract rotational energy from the black holes, is discussed in this section. The rotational energy extraction requires the presence of an ergoregion. Where, the region between an event horizon and the static limit surface (SLS) is referred to as an ergoregion. There is considerable example in which a rotating spacetime forms an ergoregion without an event horizon \cite{Glass:2004rr}. However, in this work, we have considered spacetime in which the horizon is present without a spacetime singularity.

In order to understand this process, let us assume an incident particle (0) splits into two particles ((1) and (2)) in the ergoregion. One of them, (1) crosses the event horizon while the other one, (2) escapes to infinity. As a consequence, the escaping one (2) might have energy higher than the original (0) test particle. Assume that the particle (0) falling into an ergoregion with the energy $E_{(0)} \geq 1$. As the particle will be separated into two fragments in the ergoregion, the energy will be $E_{(1)}$ and $E_{(2)}.$ As mentioned above, the particle (2) escapes to the infinity with the energy $E_{(2)} > 0,$ whereas the other one falls into the black hole with negative energy $E_{(1)} < 0.$ With this in consideration, the particle will follow the conservation laws for different parameters in the ergosphere can be written as:           
$$ E_{(0)} = E_{(1)} + E_{(2)}, $$ 
$$ L_{(0)} = L_{(1)} + L_{(2)}, $$
$$ M_{(0)} = M_{(1)} + M_{(2)}. $$
According to the formalism of Penrose process, the condition $E_{(1)} < 0$, should be fulfilled. In order to figure out how efficient energy extraction is, we consider a very simple scenario in which the test particles are restricted on an equatorial ($\theta\,=\,\pi/2$) plane, thus the conserved momentum is:
$$ P_{(0)}^{\mu} = P_{(1)}^{\mu} + P_{(2)}^{\mu}. $$
The momenta of three particles $P_{j}^{\mu}$ $(j = 0, 1, 2)$ are non spacelike and therefore it should lie inside the local light cone. The orbit of the particle moving on a plane is described by two dimensional coordinates: radial and angular coordinates (r and $\phi$). Then we can write the momentum of a particle along the geodesic $\gamma$ as,
\begin{equation}
    P_{\gamma} = P^{t} \left( \frac{\partial}{\partial t} + v \frac{\partial}{\partial r} + \Omega \frac{\partial}{\partial \phi}\right),
    \label{fourmoemntum}
\end{equation}
where, $v = dr/dt$ and $\Omega = d\phi/dt.$ The conserved energy relation is defined as $E = -P_{t}$ and it gives, 
\begin{eqnarray}
&& P^{t} = -\frac{E}{X},  \\
&& X = g_{tt} + \Omega g_{t\phi},
\label{ptandx}    
\end{eqnarray}
now from $P^{\mu}P_{\mu} = - M^2,$ we get an expression of the geodesic motion as following,
\begin{equation}
    g_{tt} + v^{2} g_{rr} + 2 \Omega g_{t\phi} + \Omega^{2} g_{\phi \phi} = - \left(\frac{M X}{E}\right)^{2}, 
    \label{geodesic}
\end{equation}
by solving the above Eq. (\ref{geodesic}) with respect to an asymptotic observer one may get the angular velocity  ($\Omega$) as,
\begin{equation}
    \Omega_{\pm} = -\frac{g_{t\phi}}{g_{\phi\phi}} \pm \sqrt{\left(\frac{g_{t\phi}}{g_{\phi\phi}}\right)^{2} - \frac{g_{tt}}{g_{\phi \phi}}},
    \label{omegapm}
\end{equation}
which denotes the angular velocity of a locally non rotating observer at a given radius r. The conservation of energy ($E = - P^{t}\, X$) and angular momentum ($L = P^{t}\, \Omega$) can be written as:
\begin{equation}
    P^{t}_{(0)} X_{(0)} = P^{t}_{(1)} X_{(1)} + P^{t}_{(2)} X_{(2)},
    \label{energy}
\end{equation}
\begin{equation}
    P^{t}_{(0)} \Omega_{(0)} = P^{t}_{(1)} \Omega_{(1)} + P^{t}_{(2)} \Omega_{(2)}.
    \label{angular}
\end{equation}
As we have mentioned, the particle (1) which crosses the horizon, will have negative energy i.e. \,$E_{(1)} < 0 $, whereas, the second particle (2) will escape to infinity as it gets rotational energy by the Penrose process. Therefore, the efficiency for the energy extraction in the Penrose process is define as,
\begin{equation}
    \eta = \xi - 1,
\end{equation}
where, 
\begin{equation}
    \xi =\frac{E_{(2)} }{E_{(0)}}.
\end{equation}
From the Eqs.\,(\ref{energy}) and (\ref{angular}), one can redefine the $\xi$ with the angular velocity as,
\begin{equation}
    \xi = \frac{\left(\Omega_{(0)} - \Omega_{(1)}\right) X_{(2)}}{\left(\Omega_{(2)} - \Omega_{(1)}\right) X_{(0)}}.
    \label{xieqn}
\end{equation}
Note that, here we consider the case, in which an incident particle has $ E_{(0)} = M_{(0)}$, and assume that it will decay into two fragments with momentum $P_{(1)}$ and $P_{(2)}$. Now, from Eq.\,(\ref{xieqn}), one can see that the efficiency $\eta$ = $ \xi -$ $1$ is maximized when we consider the largest value of $\Omega_{(2)}$ and the smallest value of $\Omega_{(1)}$. We can get maximum efficiency when the term $dr/dt$ vanishes in Eq.\,(\ref{fourmoemntum}) for both particles. In that case, we find:
\begin{equation}
   P_{(1)} = P^{t}_{(1)} \left( \frac{\partial}{\partial t} + \Omega_{(1)} \frac{\partial}{\partial \phi}\right),   
\end{equation}
\begin{equation}
   P_{(2)} = P^{t}_{(2)} \left( \frac{\partial}{\partial t} + \Omega_{(2)} \frac{\partial}{\partial \phi}\right). 
\end{equation}
and then using the Eqs. (\ref{ptandx}) and (\ref{geodesic}) we can get the expression of $  \Omega_{(0)} $ as:
\begin{equation}
    \Omega_{(0)} = \frac{- g_{t\phi} (1 + g_{tt}) + \sqrt{(1 + g_{tt})(g_{t\phi}^{2} - g_{tt} g_{t\phi} )}}{ g_{t\phi}^{2} + g_{\phi \phi}}.
    \label{omegazero}
\end{equation}
\\
So ultimately, from the Eqs.(\ref{omegapm}), (\ref{xieqn}) and (\ref{omegazero}), we can obtain the general expression of the efficiency rate for the maximum extracted energy as:
\begin{equation}
\eta_{max} \leq \frac{g_{\phi \phi} (\sqrt{1 + g_{tt}} + 1) + g_{t\phi}^{2}}{2 g_{\phi \phi}\sqrt{1 + g_{tt}}}   - 1. 
\label{engeffi}
\end{equation}
The maximum energy can be extracted if we consider the fragment splitting at the outer horizon and the same scenario is given for Penrose process. In expression (\ref{engeffi}), an equality shows that the splitting of a particle is happening at the outer horizon. Now, if we use the metric tensor components of the Kerr black hole in Eq.\,(\ref{engeffi}), then we can get the maximum efficiency of extracted energy for extreme spin parameter ($a = M$) is $20.7 \%$. Now, in the next section, we will look for rotating Simpson-Visser spacetime for the same.
\section{Rotating Simpson Visser Spacetime}
\label{SVsection}

In \cite{Simpson:2018tsi}, Simpson and Visser proposed a spherically symmetric spacetime that smoothly interpolates between a Schwarzschild black hole ($l\,=\,0$) and Morris- Thorne wormholes with regular geometry. However, the more physical scenario can be considered by introducing the spin parameter ($a$) in this metric. The rotating Simpson-Visser spacetime is derived using the Janis-Newmann algorithm in \cite{Shaikh:2021yux,Mazza:2021rgq}. The metric for that spacetime can be written as:
\begin{widetext}
\begin{eqnarray}
  ds^{2} = -\left(1 - \frac{2 M \sqrt{r^2 + l^2}}{A}\right) dt^{2}  + \frac{A}{\Delta} dr^{2} + A d\theta^2  - \frac{4 M a \sqrt{r^2 + l^2}  sin^{2}\theta}{A} dt d\phi  \nonumber \\ +\left(r^2 + a^2 + l^2+\frac{2 M a^2 \sqrt{r^2 + l^2}  sin^{2}\theta}{A}\right) sin^{2}\theta d\phi^{2} ,
  \label{rotatingSV}
\end{eqnarray}
\end{widetext}
where, $$ A = r^2 + l^2 + a^2 cos^{2}\theta, \hspace{0.5cm} \Delta = r^2 + l^2 + a^2 - 2 M \sqrt{r^2 + l^2}.  $$
Here, $M$ denotes the ADM mass of the spacetime metric and $l$ is a regularisation parameter. The metric given in (\ref{rotatingSV}) reduces to the Kerr spacetime with $l=0$ and to  a Schwarzschild spacetime with the vanishing spin $(a)$ and regularisation $(l)$ parameters. The rotating Simpson-Visser spacetime also possesses an inner horizon and an outer horizon, as well as an ergoregion. Inner and outer horizons are known as the Cauchy horizon and an event horizon respectively. Depending on the different values of spin parameter $a$ and a regularisation parameter $l$, the nature of the compact object changes. One can understand how these properties of the compact object described by the rotating Simpson-Visser spacetime metric changes by understanding how horizons and ergoregion are defined from the metric. For any general spacetime metric, the horizon can be defined by $g_{tt}=0$. Thus for rotating Simpson-Visser metric, one can write the expression of horizon radius as,
\begin{equation}
    r_{\pm} = \left( \left(M \pm \sqrt{M^2 - a^2}\right)^2 - l^2 \right)^{\frac{1}{2}},
    \label{ehrotating}
\end{equation}
where, $r_{+}$ and $r_{-}$ corresponds to outer and inner horizons respectively.

Since we have considered a rotating black hole, the spacetime region around the center of the compact object also possesses rotational motion. This effect is known as the frame dragging effect. In the spacetime region upto a certain radius, frame dragging effect is so prominent that all particles also rotate with the rotating spacetime region around the compact object. The spacetime region where this effect is observed is known as an ergoregion. An observer can never be stationary in this region. Depending on values of different parameter in the spacetime metric components, the ergoregion changes. To study the change in the ergoregion, we need to know the mathematical expression of the ergoregion. For any general spacetime metric, the ergoregion can be defined by $(g_{rr})^{-1}\,=\,0$. From which the radius of the ergosphere in rotating Simpson Visser can be expressed as,
\begin{equation}
    r_{erg\pm}^{2} = \left(M \pm \sqrt{\left(M^2 - a^2 cos^{2}\theta \right)}\right)^2 - l^2.
    \label{rotatingergo}
\end{equation}

Rotating Simpson-Visser spacetime suggests a regular geometry as $l$ is always positive ($l \neq 0$). Thus singularity would not exist for the rotating Simpson-Visser even at $r=0$ and the metric represents a finite size surface with a regular geometry.

From Eqs.\,(\ref{ehrotating}) and (\ref{rotatingergo}), one can see that the mathematical expressions of those equations would be imaginary for certain values of $a$, $M$ and $l$. Thus depending on different values of these quantities, physical properties of horizons and the ergoregion change and thus the nature of the compact object also changes. For these equations, their expressions are mathematically real and thus physical only when $a<M$. For $a>M$, horizons would not exist as the expressions of Eqs.\,(\ref{ehrotating}) and (\ref{rotatingergo}) become imaginary. Such a geometry represents a wormhole. While for $a<M$, the shape of the ergoregion and the existence of horizons depend on the regularisation parameter $l$. If the regularisation parameter $l$ is less than $M+\sqrt{M^2-a^2}$ in the Eq.\,(\ref{ehrotating}), then the event horizon exists. While for existence of the Cauchy horizon, the condition $l<M-\sqrt{M^2-a^2}$ needs to be fulfilled. In such cases, compact object would have both Cauchy horizon and event horizon with an ergoregion around them. That kind of geometry is known as Regular Black Hole- 2. But for $a<M$, if the condition $l<M+\sqrt{M^2-a^2}$ is satisfied but value of regularisation parameter $l$ is larger than $M-\sqrt{M^2-a^2}$ then only an event horizon would exist as the expression of $r_{-}$ becomes imaginary in Eq.\,(\ref{ehrotating}). This type of compact object is known as Regular Black Hole-1. However, for $l>M+\sqrt{M^2-a^2}$, the geometry would not possess any horizon and thus it cannot be termed as a regular black hole geometry. The different geometries of rotating Simpson-Visser spacetime related to different values of spin parameter and regularisation parameter is consistent with \cite{Islam:2021ful}.

\begin{figure*}
\centering
\subfigure[ Regular Black hole-2, a=0.1, l=0.005]
{{\includegraphics[width=5cm]{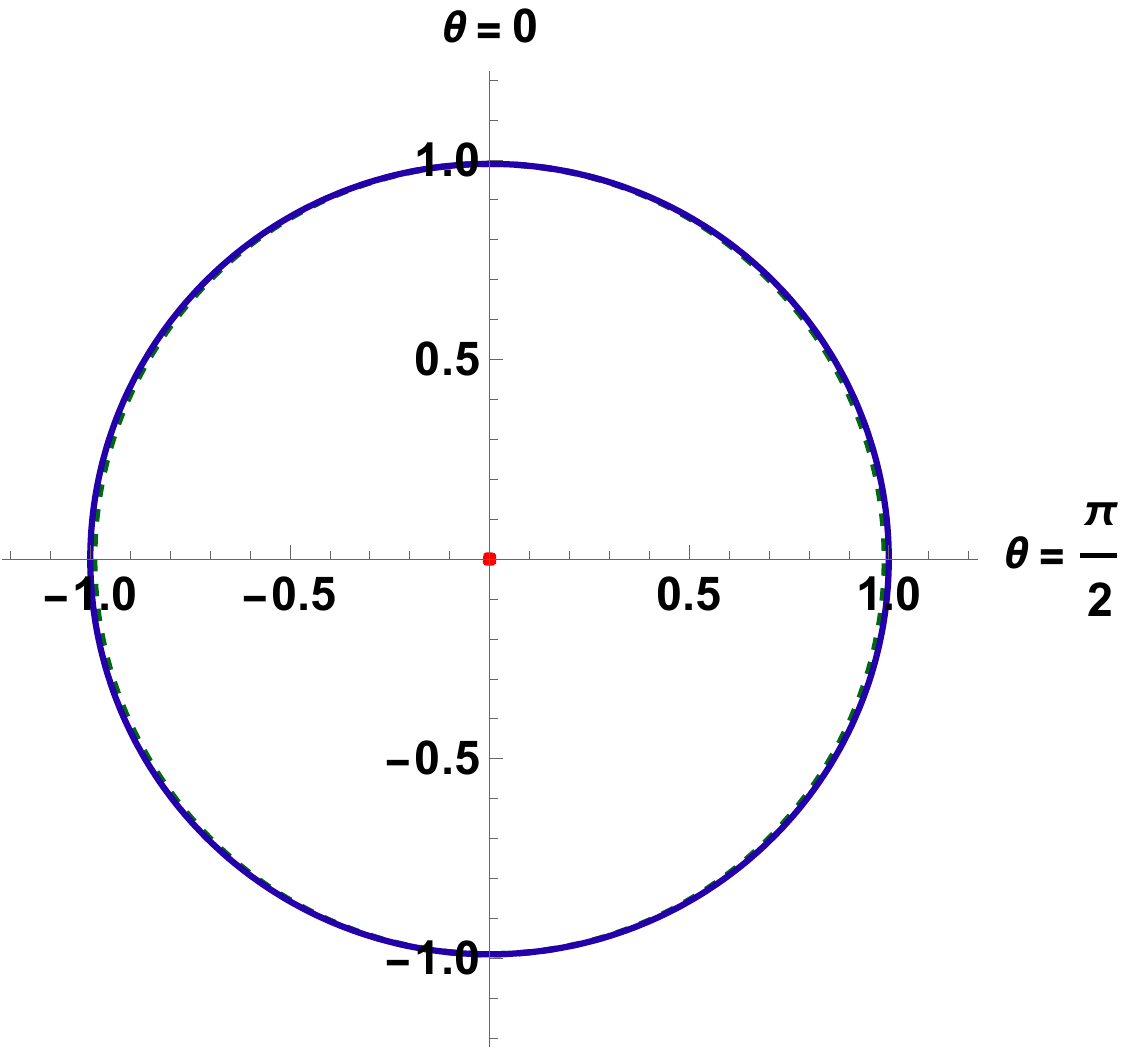}}
\label{11}}
\hspace{0.2cm}
\subfigure[ Regular Black hole-1, a=0.1, l=0.4]
{{\includegraphics[width=5cm]{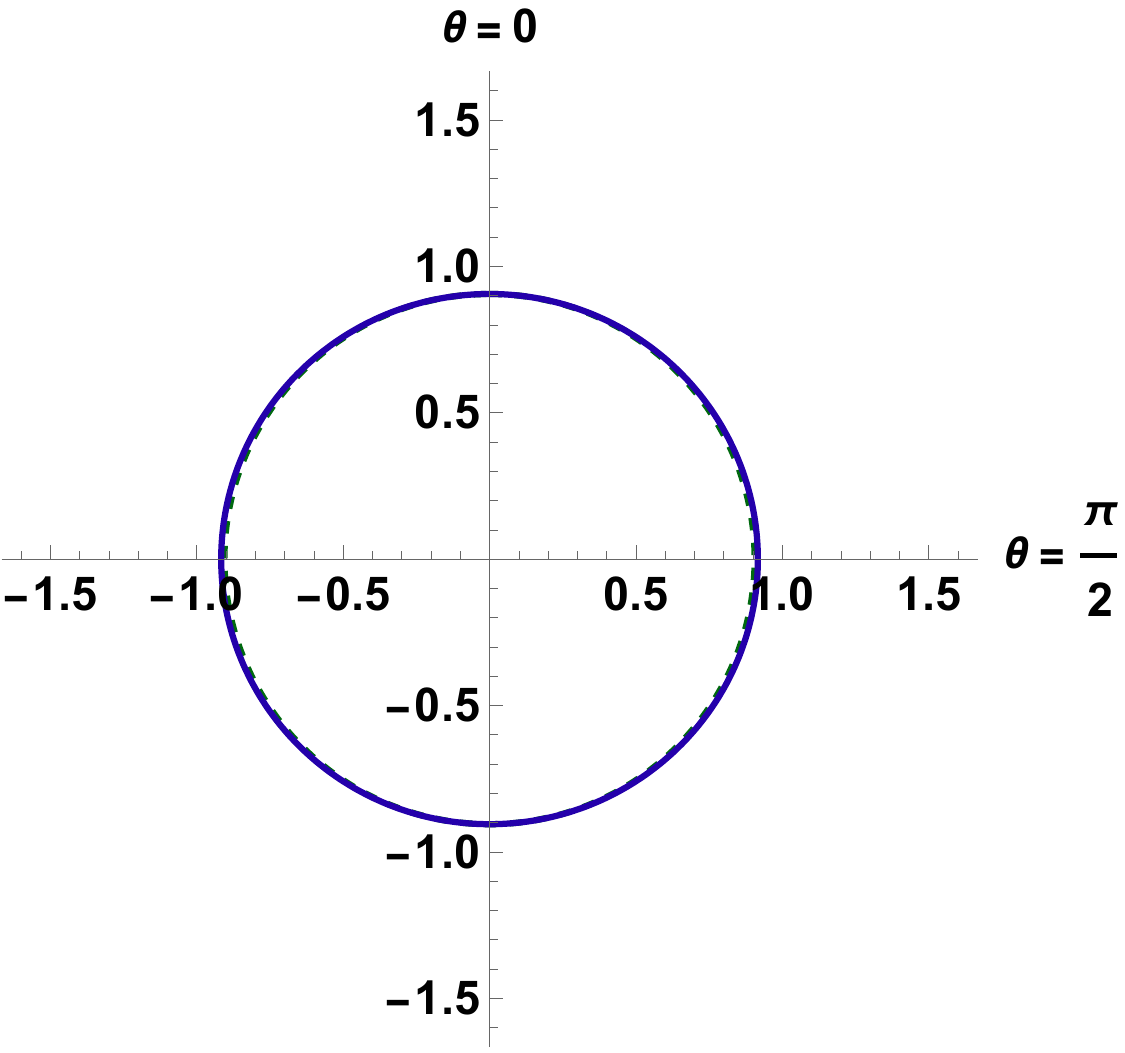}}
\label{22}}
\hspace{0.2cm}
\subfigure[ Wormhole, a=0.1, l=0.99]
{{\includegraphics[width=5cm]{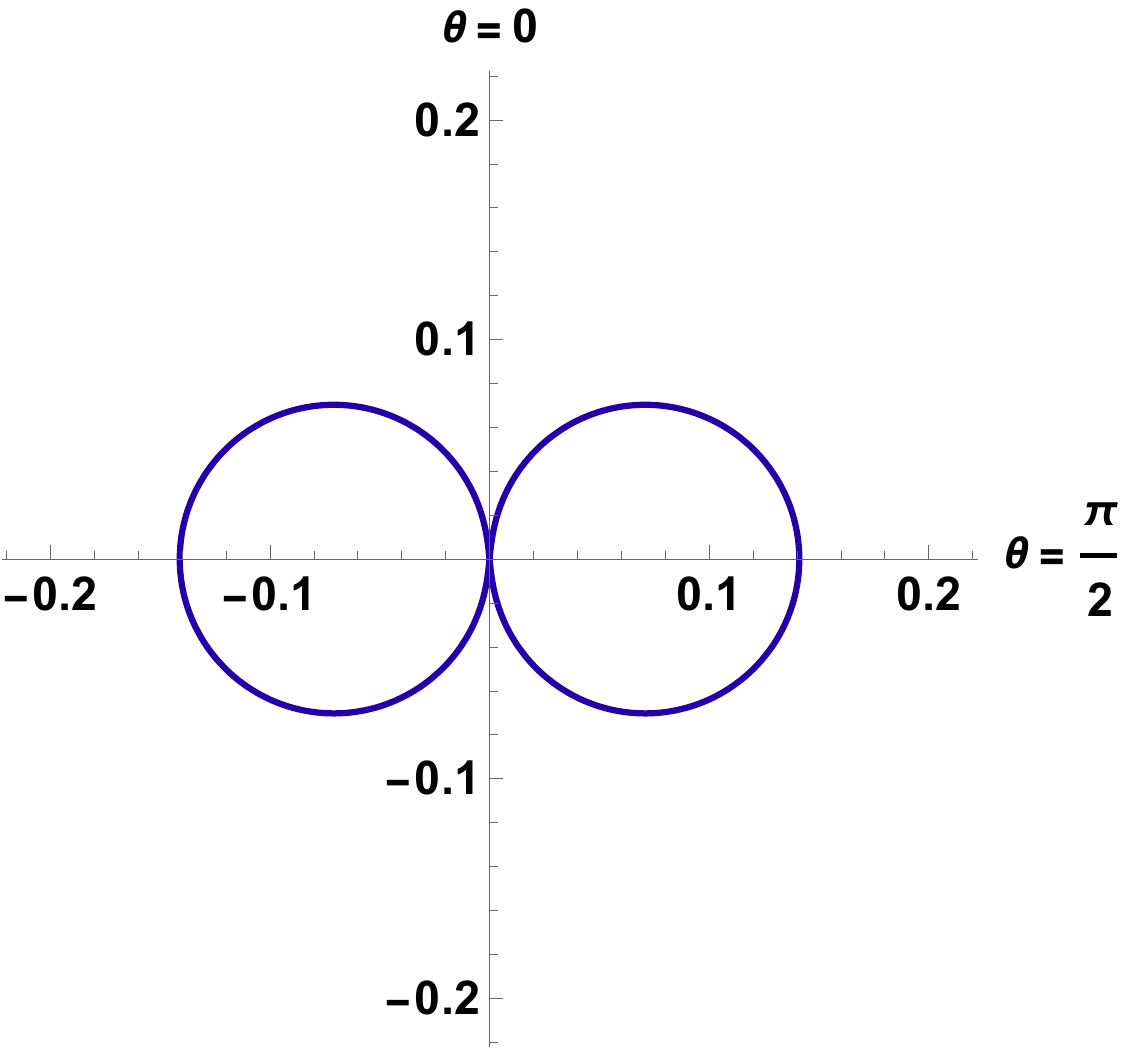}}
\label{33}}
\hspace{0.2cm}   
\subfigure[ Regular Black hole-2, a=0.2, l=0.01]
{{\includegraphics[width=5cm]{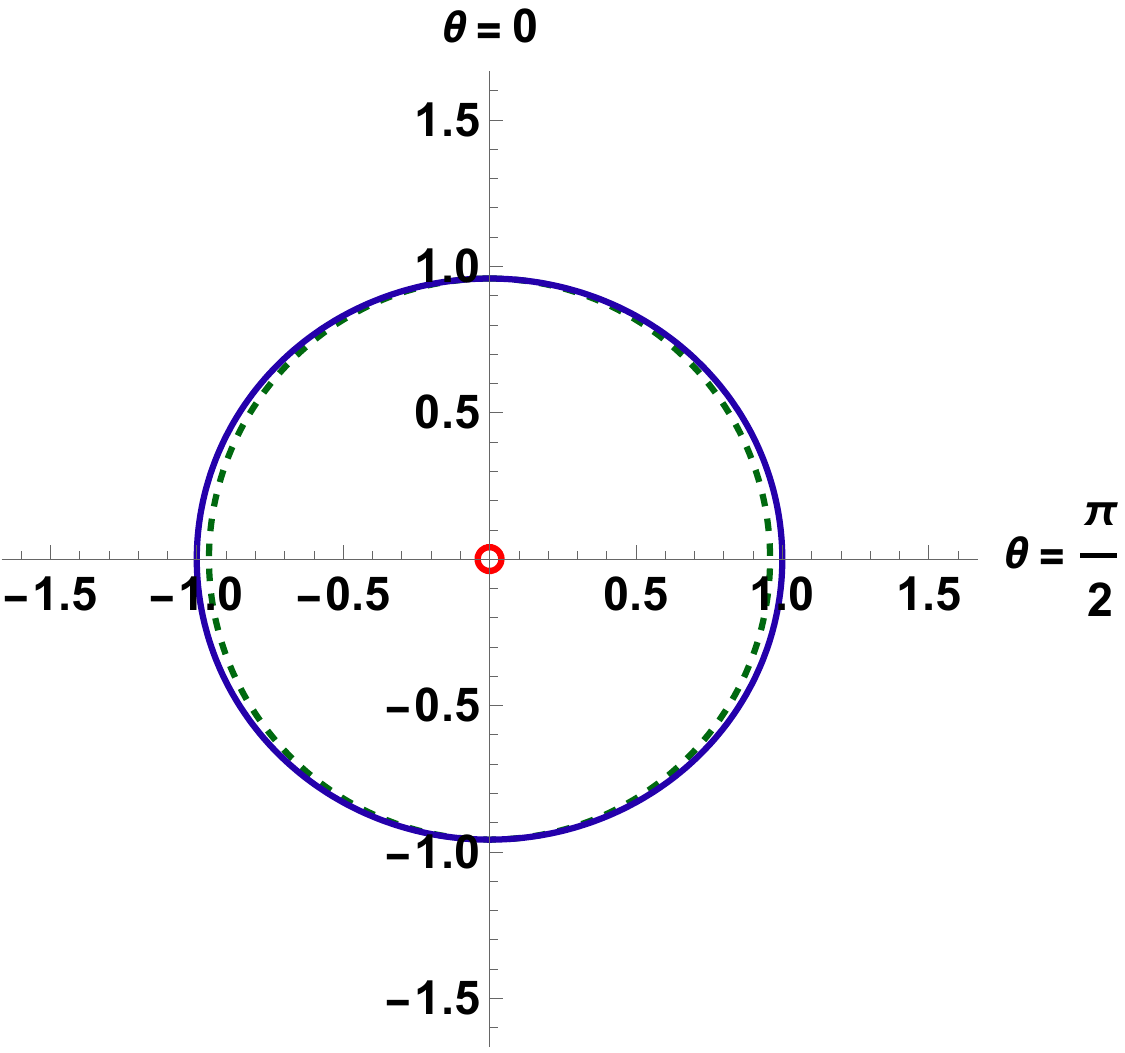}}
\label{a=0.2, l=0.01}}
\hspace{0.2cm}
\subfigure[ Regular Black hole-1, a=0.2, l=0.5]
{{\includegraphics[width=5cm]{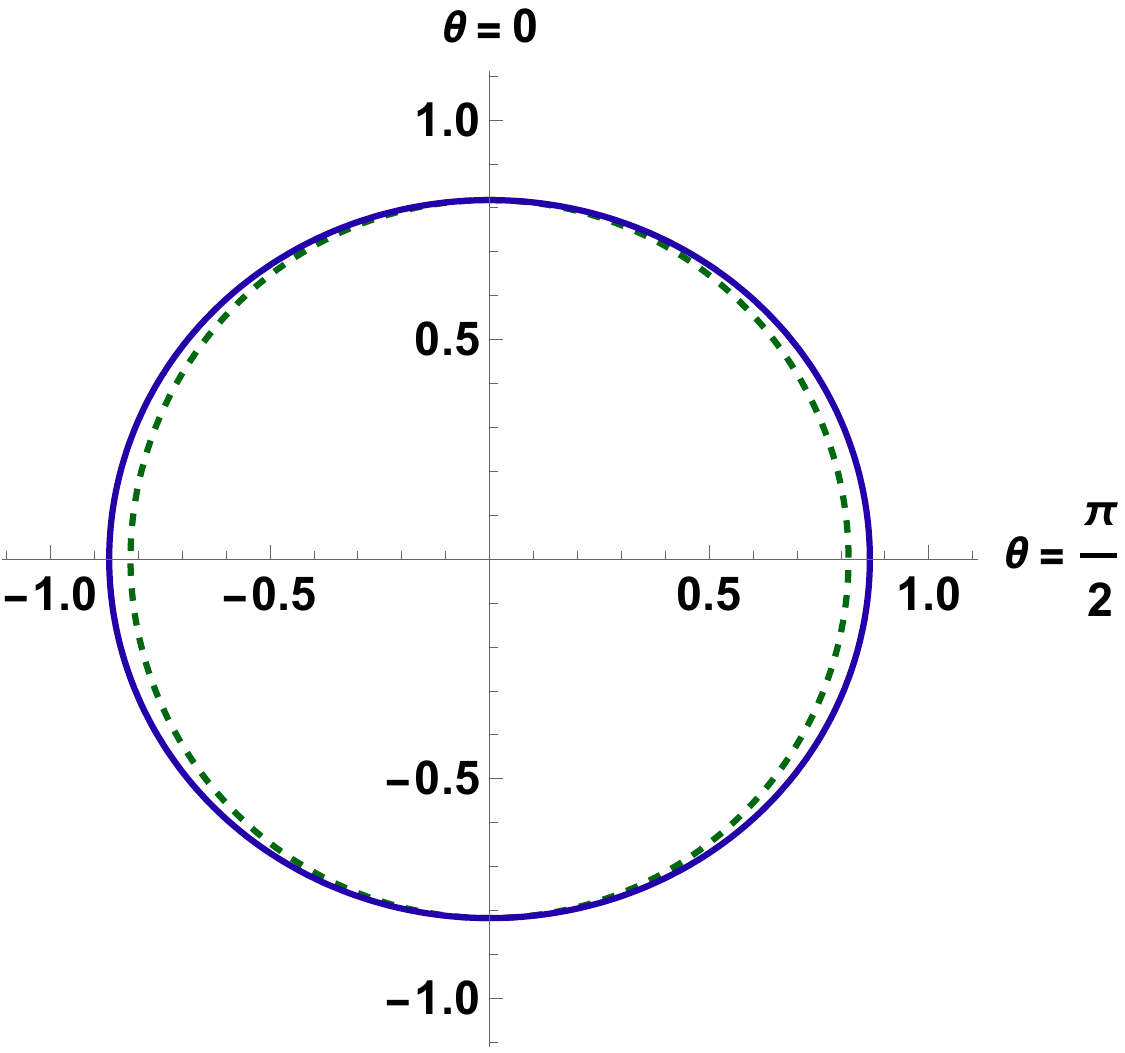}}
\label{a=0.2, l=0.5}}
\hspace{0.2cm}
\subfigure[ Regular Black hole-2, a=0.3, l=0.08]
{{\includegraphics[width=5cm]{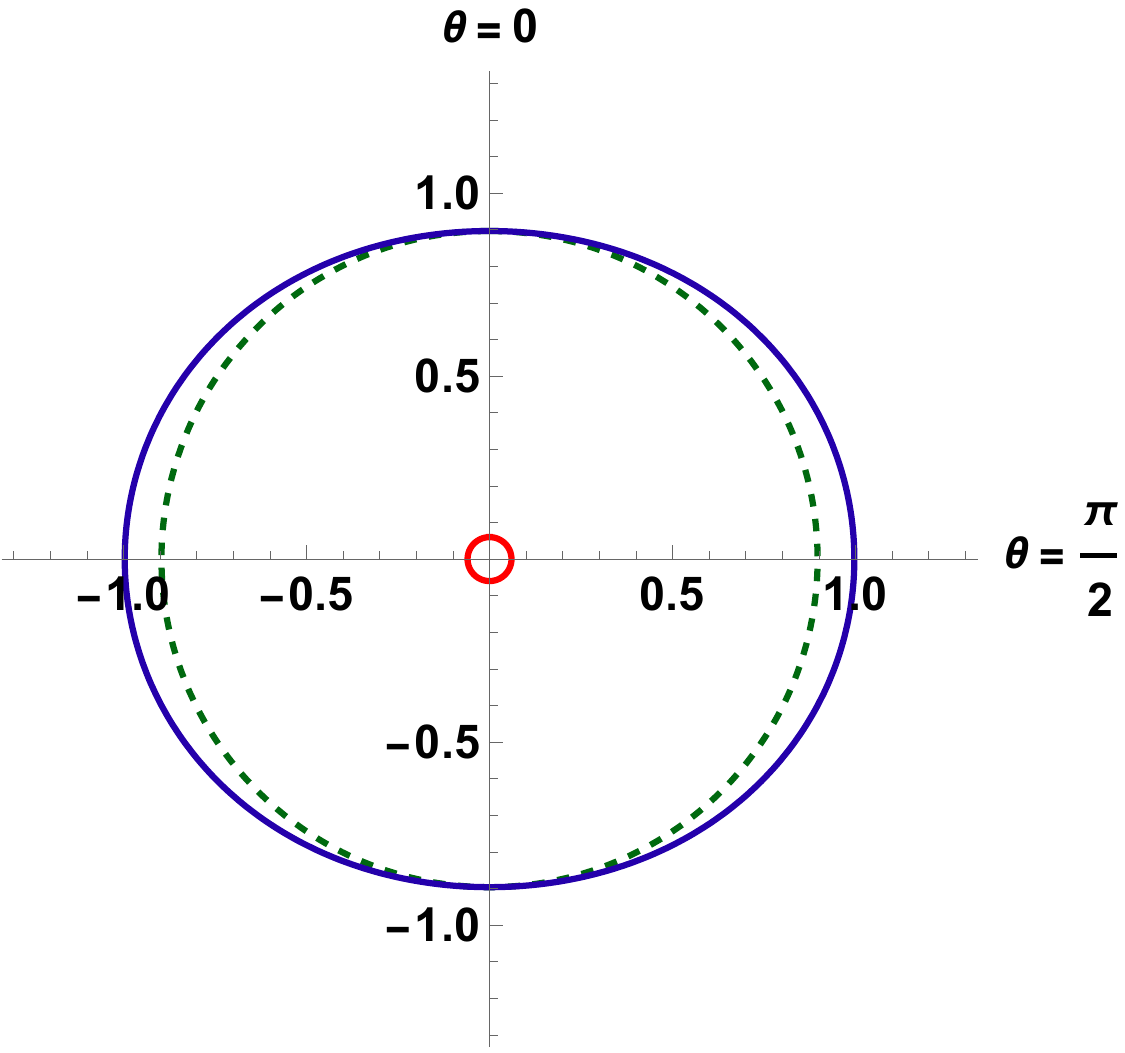}}
\label{a=0.3, l=0.08}}
\hspace{0.2cm}
\subfigure[ Regular Black hole-1, a=0.3, l=0.5]
{{\includegraphics[width=5cm]{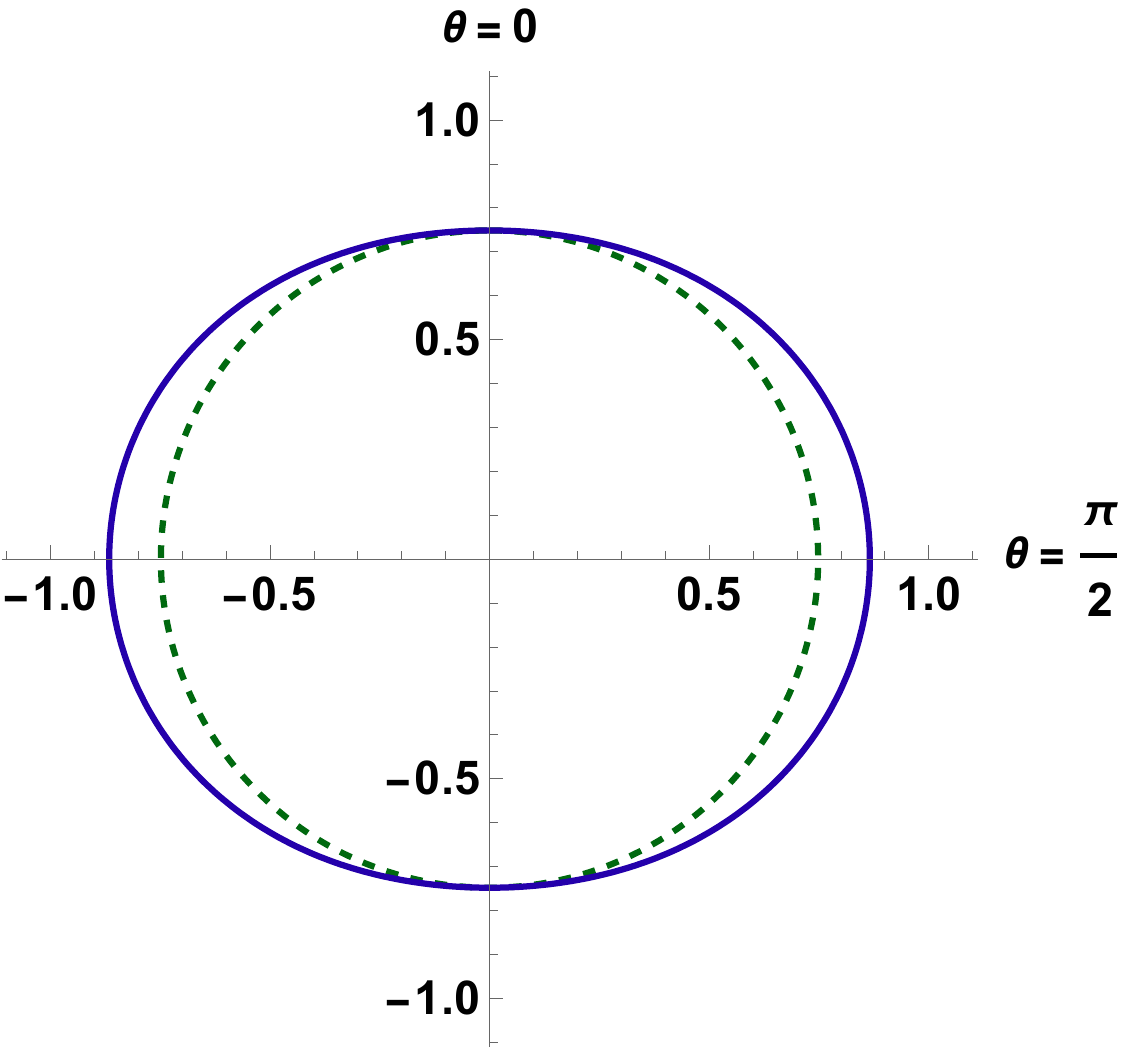}}
\label{a=0.3, l=0.5}}
\hspace{0.2cm}
\subfigure[ Regular Black hole-2, a=0.4, l=0.1]
{{\includegraphics[width=5cm]{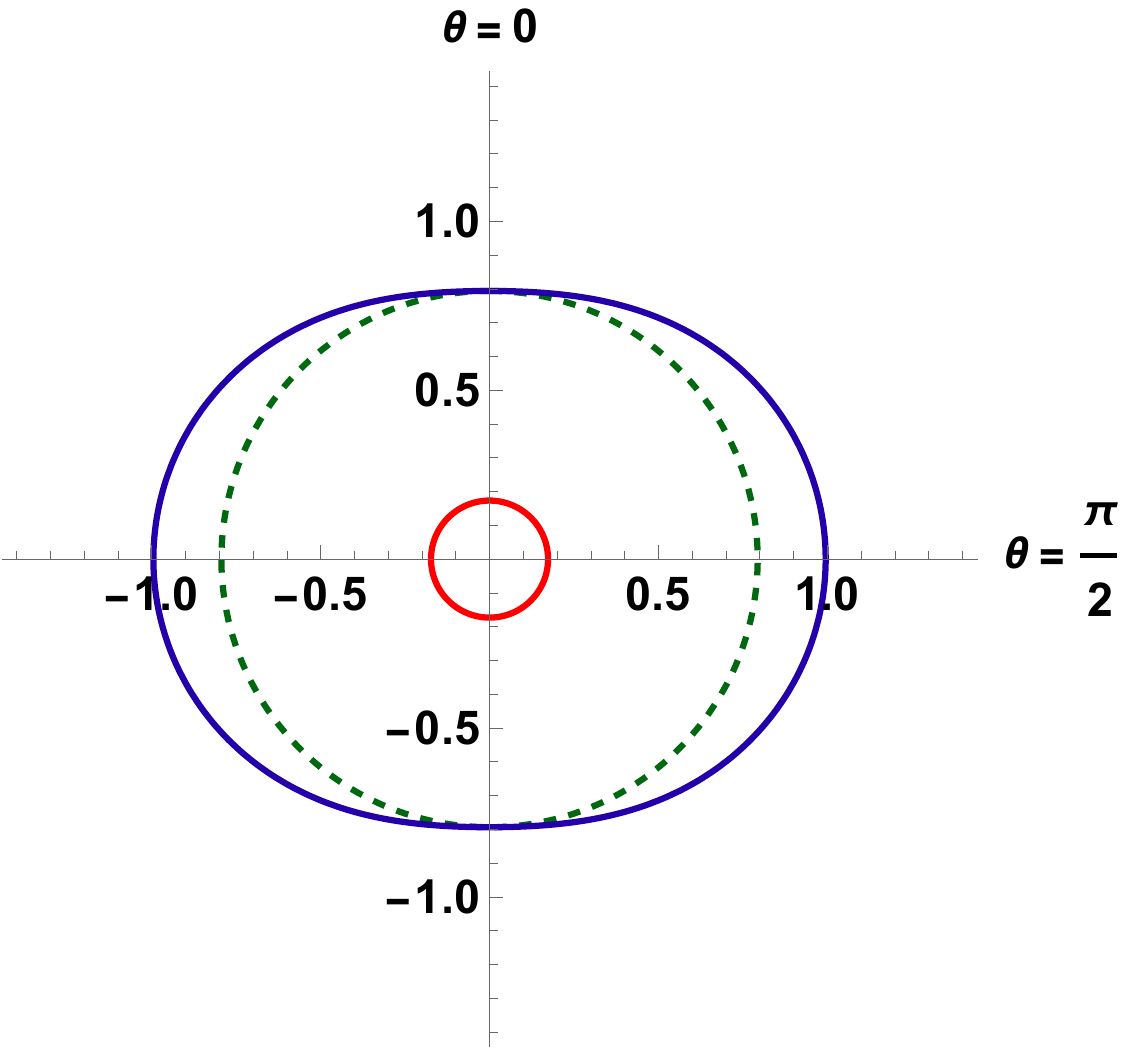}}
\label{a=0.4, l=0.1}}
\hspace{0.2cm}
\subfigure[ Regular Black hole-1, a=0.4, l=0.4]
{{\includegraphics[width=5cm]{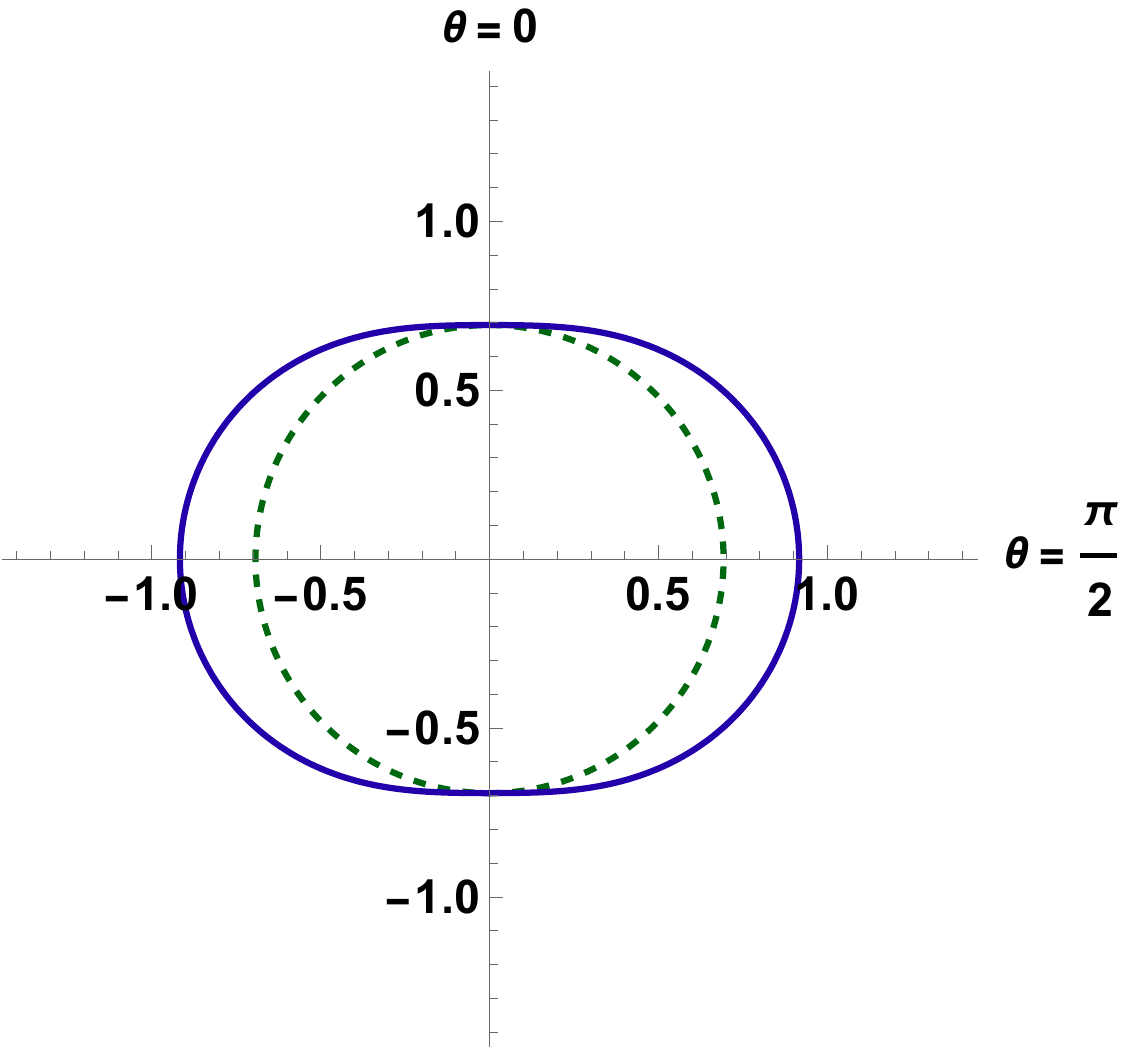}}
\label{a=0.4, l=0.4}}
\hspace{0.2cm}
\subfigure[ Extremal regular black hole, a=0.5, l=0.3]
{{\includegraphics[width=5cm]{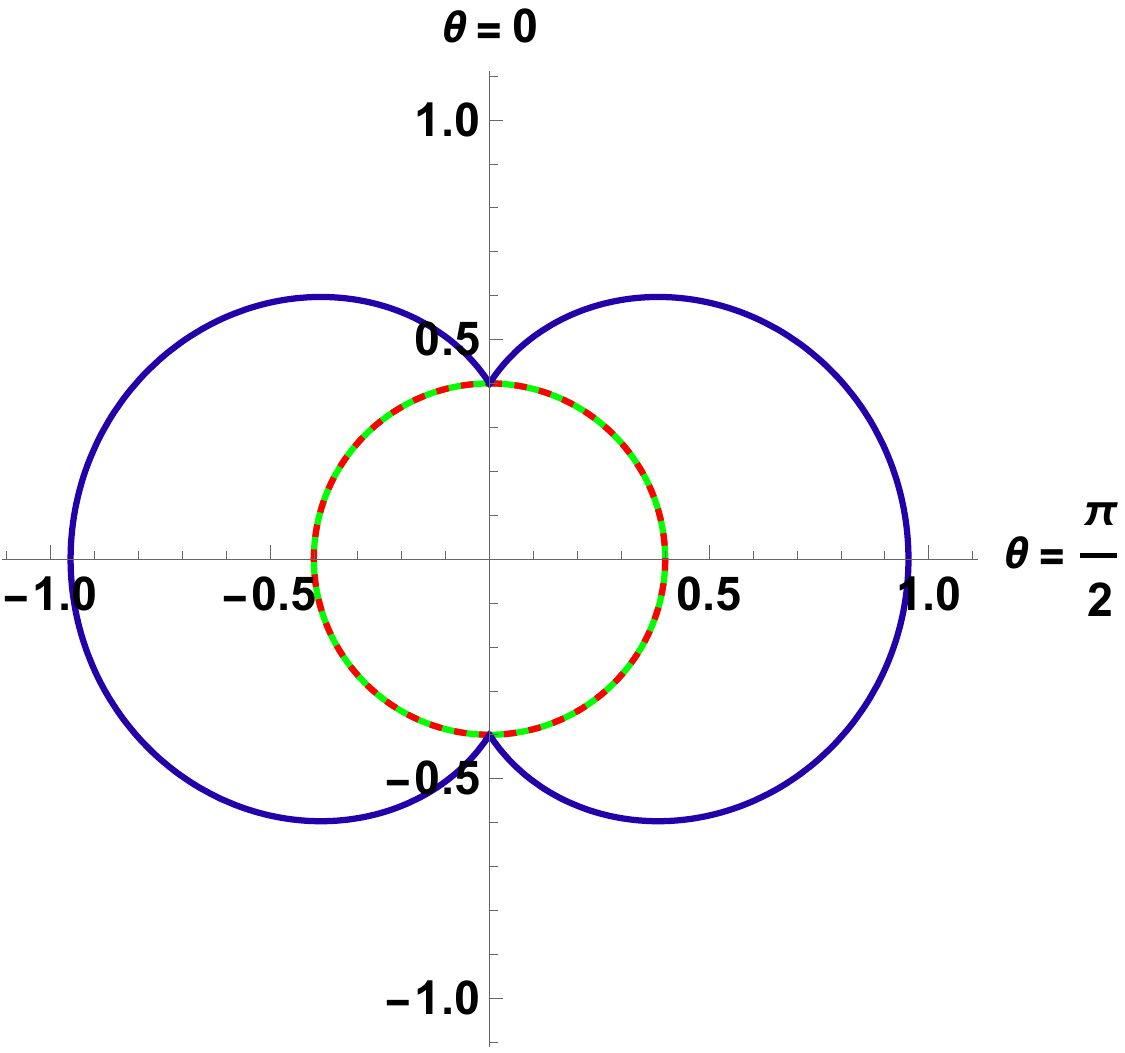}}
\label{ a=0.5, l=0.3}}
\hspace{0.2cm}
\subfigure[ Wormhole, a=0.5, l=0.7]
{{\includegraphics[width=5cm]{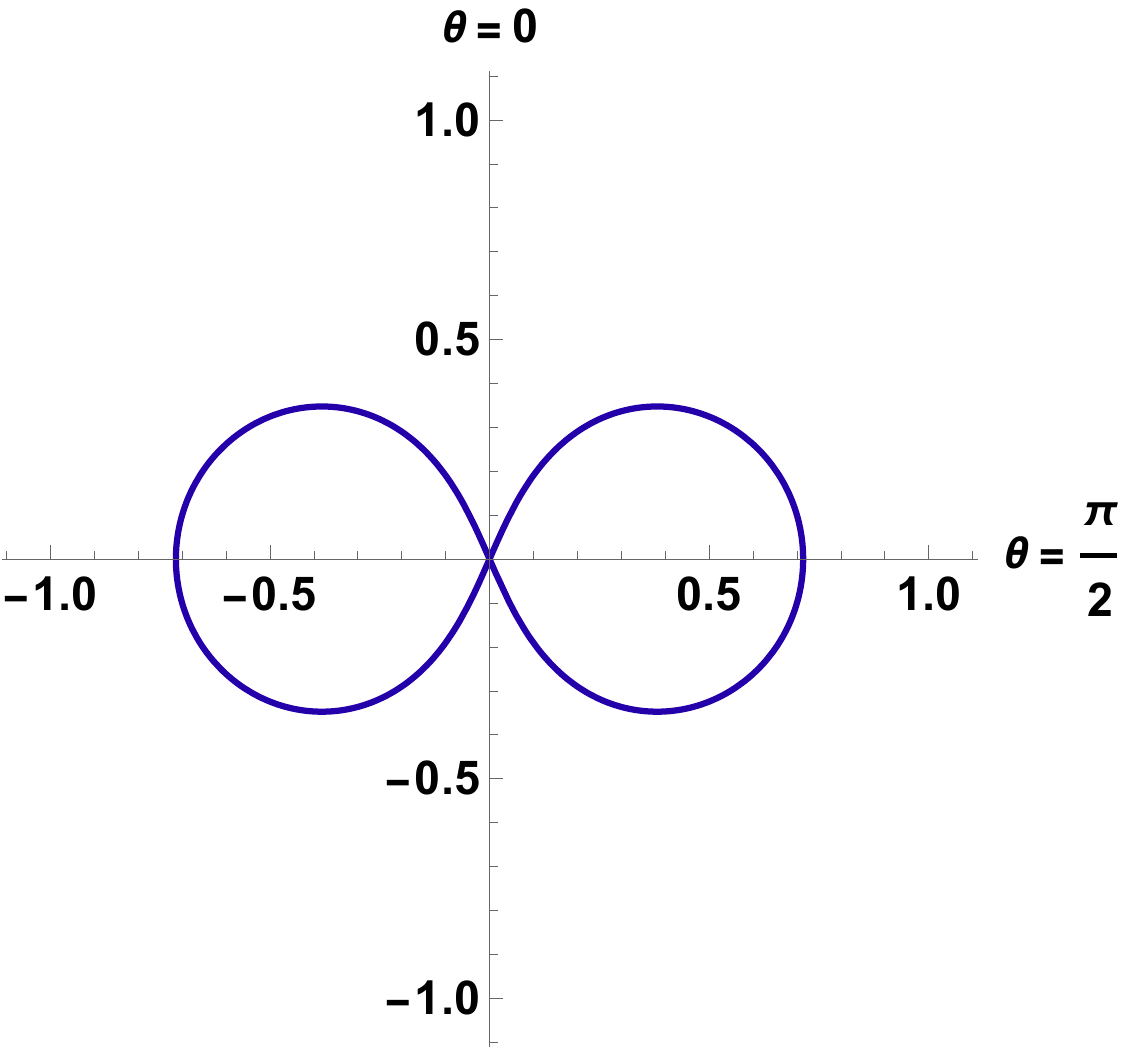}}
\label{ a=0.5, l=0.7}}
\hspace{0.2cm}
\subfigure[a=0.6, l=0.3]
{{\includegraphics[width=5cm]{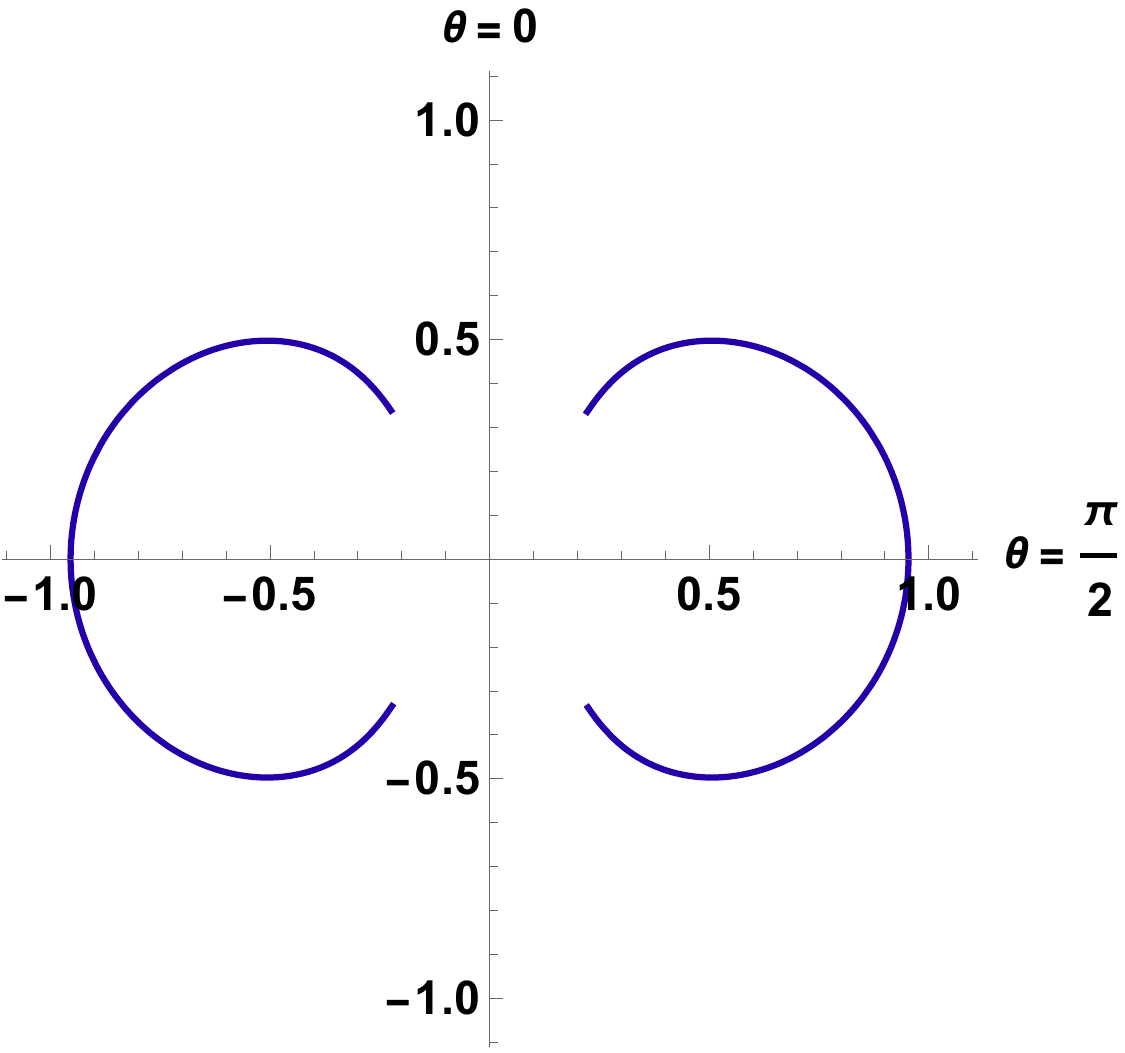}}
\label{ a=0.6, l=0.3}}
\hspace{0.2cm}

 \caption{Figures shows the behaviour of ergoregion and inner/outer horizon in the rotating Simpson-Visser Spacetime with different spin and regularisation parameters. The blue circle represent the boundary of ergoregion and the green dotted circle represents the event horizon and the red line represents inner horizon. Where, we consider $M=0.5$.}
  \label{SVEHplot}
\end{figure*}

Till now we have discussed the cases for which $a<M$. But compact object has interesting geometrical features when $a=M$. Because in this condition, for $l<M$ the Eq.\,(\ref{ehrotating}) has same mathematical expression for Cauchy horizon radius $r_{-}$ and event horizon radius $r_{+}$. Thus both horizons would exist at just one particular radius. This compact object is termed as an extremal regular black hole with degenerate horizons. While for $a=M$ and $l>M$, the geometry would not possess any horizon as the expression in the Eq.\,(\ref{ehrotating}) becomes imaginary and thus the compact object is not a black hole but rather a wormhole. One should note that, as the spin parameter $a$ is increased, the area of the ergoregion also increases as one can see in Figs.\,(\ref{SVEHplot}).

For the given rotating Simpson-Visser metric, we have shown different shapes ergoregions corresponding to different values of $a$ and $l$ in \ref{SVEHplot}. Starting with the low spin parameter $a=0.1$, we get Regular Black Hole- 2 with both inner and outer horizons for value of regularisation parameter $l=0.005$ which is less than $0.5-\sqrt{0.5^2-0.1^2}$ as shown in Fig.\,(\ref{11}). While for value of regularisation parameter $0.5-\sqrt{0.5^2-0.1^2}<l<0.5+\sqrt{0.5^2-0.1^2}$, there exist just event horizon with Regular Black Hole-1 geometry as one can see in Fig.\,(\ref{22}). Finally, when $l>M+\sqrt{M^2-a^2}$, expressions of both event horizon and Cauchy horizon becomes imaginary Fig.\,(\ref{33}). In the similar pattern, we have shown plots for different values of $l$ which corresponds compact objects with different physical and geometrical properties as we increase the value of spin parameter $a$ keeping the mass of the compact object $M=0.5$. We get similar plots till we increase the spin parameter upto $a<M$ as one can see in Fig.\,\ref{SVEHplot}.  One should note that, as spin parameter $a$ increase, the ergoregion also becomes larger which can be seen in Fig.\,\ref{SVEHplot}. 

However, when we change the spin parameter to $a=0.5$, which is similar to the value of the mass of the compact object $M=0.5$, we get different scenarios for different values of $l$. For $a=0.5$, when $l<a$ Cauchy horizon radius $r_{-}$ and event horizon radius $r_{+}$ has the same value. An extremal black hole representing this geometry with degenerate horizons can be seen in Fig.\,\ref{ a=0.5, l=0.3}. While for same spin parameter $a=0.5$, if the value of the regularisation parameter $l$ is larger than the value of the spin parameter $a=0.5$, then the compact object does not possess any horizon as the expression of horizon in Eq.\,(\ref{ehrotating}) again becomes imaginary and the object is a wormhole as one can see in Fig.\,\ref{ a=0.5, l=0.7}. Now finally considering the case where the spin parameter is taken to be $a=0.6$, which is larger than $M=0.5$, the expression of Cauchy horizon, event horizon and ergoregion becomes imaginary which can be seen in Fig.\,\ref{ a=0.6, l=0.3}. This geometry again represents a wormhole. One can go into the details of these different geometries, especially nature of the throat of the wormholes for different values of regularisation parameter $l$ and spin parameter $a$ by studying their corresponding Penrose diagrams given in the \cite{Mazza:2021rgq}.

\subsection{Energy Extraction by Penrose process from rotating Simpson Visser spacetime}

As we have mentioned earlier, the extraction of energy by Penrose process from rotating object, the existence of the ergoregion and horizon is important. In the previous section, we discussed how ergoregion and horizons are changing with different values of the spin parameter and the regularisation parameter.  It is evident that the expressions of horizons and ergoregion are different from those of a Kerr black hole as one can see in Eq.\,(\ref{ehrotating}) and Eq.\,(\ref{rotatingergo}). As a consequence, energy extraction efficiency should differ from the Kerr black hole. To check the energy extraction efficiency in the rotating Simpson-Visser spacetime, the angular velocity with respect to an asymptotic observer at the infinity from Eq.\,(\ref{omegapm}) can be written as,
\begin{widetext}
\begin{eqnarray}
\Omega{\pm} =  \sqrt{\frac{\csc ^2(\theta )
   \left(a^2-2 M \sqrt{l^2+r^2}+l^2+r^2\right) \left(a^2 \cos (2 \theta )+a^2+2 \left(l^2+r^2\right)\right)^2}{\left(a^4+a^2 \cos (2 \theta ) \left(a^2-2 M
   \sqrt{l^2+r^2}+l^2+r^2\right)+a^2 \left(2 M \sqrt{l^2+r^2}+3 l^2+3 r^2\right)+2 \left(l^2+r^2\right)^2\right)^2}} \nonumber\\
   \pm \frac{2 a M \sqrt{l^2+r^2}}{\left(a^2+l^2+r^2\right)^2-a^2 \sin ^2(\theta ) \left(a^2-2 M \sqrt{l^2+r^2}+l^2+r^2\right)}.
\end{eqnarray}
\end{widetext}

The above expression shows the angular velocity with respect to an asymptotic observer which resides on an equatorial plane. The maximum change on an ergoregion  due to the rotation can be observed at $ \theta\,=\,\pi/2$. Therefore, the maximum energy efficiency which can be extracted from rotating Simpson-Visser spacetime is derived using (\ref{engeffi}),
\begin{widetext}
\begin{eqnarray}
&& \eta_{max} = \frac{1}{4}  \left(\frac{2 \sqrt{2} a^2 M}{\left( \sqrt{\frac{M}{\sqrt{2 M \left(\sqrt{M^2-a^2}+M\right)-a^2}}} \right)f_{1}(M,a)}  + h_{1}(M, a) -2\right) 100,
\end{eqnarray}
\end{widetext}
where $ f_{1}(M,a) $ and $h_{1}(M, a) $ are,
\begin{widetext}
\begin{eqnarray}
&& f_{1}(M,a) =\left(a^2 \left(-\sqrt{M^2-a^2}+\sqrt{2 M \left(\sqrt{M^2-a^2}+M\right)-a^2}-3 M\right) + 4 M^2 \left(\sqrt{M^2-a^2}+M\right)\right),  \\
&&   h_{1}(M, a) = \frac{\sqrt{2}}{\sqrt{\frac{M}{\sqrt{2 M \left(\sqrt{M^2-a^2}+M\right)-a^2}}}}.
\end{eqnarray}
\end{widetext}
Surprisingly, it can be seen in the above expression that the energy efficiency depends only on the spin parameter $a$ and not on the regularisation parameter $l$. In table (\ref{tabsv}),\, energy extraction efficiency for different values of $a$ and $l$ is shown. As the spin increases, the angular velocity increases, and energy extraction efficiency also increases. For $l=0$, we get the energy efficiency for the Kerr metric. With the constant spin parameter, energy efficiency is the same for different values of the regularisation parameter $l$ as the energy efficiency does not depend on it. Thus for regular black holes, energy extraction efficiency is the same as it is in the Kerr black hole case. For some cases, energy extraction by the Penrose process is not possible in rotating Simpson-Visser spacetime as horizon does not exist in certain conditions as discussed previously. As discussed in the Section (\ref{PP}), it is also important to point it out that, energy extraction for Penrose process is defined for the case in which particle splitting occurs at the event horizon as we get the maximum efficiency in that condition. Thus for rotating Simpson-Visser metric also, Penrose process is considered to be taking place at the event horizon only in this paper. As we move away from the event horizon, the energy efficiency decreases gradually in the ergoregion and outside the ergoregion, energy extraction efficiency drops significantly. Thus we have defined Penrose process for rotating Simpson-Visser spacetime, a class of regular compact objects where singularity is absent. In the next section we discus the energy extraction in conformally transformed spacetimes.

\begin{table}[]
    \centering \scalebox{0.94}{%
    \begin{tabular}{||c c c c c c c c c c ||} 
 \hline
 No & Spin Parameter (a) &  l = 0  & l = 0.6 & l = 1.4 & l = 1.8\\ [0.5ex] 
  \hline
 1 & 0.1 & 0.0627 & 0.0627 & 0.0627 & 0.0627\\ 
 \hline
 2 & 0.2 & 0.2544 & 0.2544 & 0.2544 & 0.2544\\
 \hline
 3 & 0.3 & 0.5859 &  0.5859 & 0.5859 & 0.5859 \\
 \hline
 4 & 0.4 & 1.0774 & 1.0774 & 1.0774 & 1.0774\\
 \hline
 5 & 0.5 & 1.7638 & 1.7638 &  1.7638 & 1.7638\\ 
 \hline
 6 &  0.6 & 2.7046 & 2.7046 & 2.7046 & 2.7046\\
 \hline
 7 & 0.7 & 4.0084 & 4.0084 & 4.0084 & -\\ 
 \hline
 8 & 0.8 & 5.9017 & 5.9017 & 5.9017 & -\\ 
 \hline
 9 & 0.9 & 9.0098 & 9.0098 & 9.0098 & -\\ 
 \hline
 10 & 0.93 & 10.4661 & 10.4661 & - & -\\
 \hline 
 11 & 0.96 & 12.5 & 12.5 & - & - \\
 \hline
 12 & 0.99 & 16.1956  & 16.1956 & - & - \\
 \hline
 13 &  1   & 20.7107  & -  & - & -  \\ 
 \hline
\end{tabular}}
     \caption{Efficiency of energy extraction in the rotating Simpson-Visser spacetime for different values of $l(0, 0.6,1.4,1.8)$.}
    \label{tabsv}
\end{table}

\section{A regular and singular black hole spacetimes}
\label{regusingusection}
In \cite{Bambi:2016wdn}, a conformally transformed rotating black hole solutions are proposed. Aforementioned, these conformally transformed black hole spacetimes are solution of CEFE. In Boyer–Lindquist coordinates, the metric can be written as, 
\begin{equation}
    dS^{2} = S\,  ds^{2}_{Kerr}, 
    \label{nuvalueeqn}
\end{equation}
where,
\begin{equation}
    S = \left(1 + \frac{l^{2}}{\Sigma}\right)^{\nu}, 
    \label{rescalingfac}
\end{equation}
\hspace{0.2cm} $$ \Sigma = r^2 +  a^2 cos^{2}\theta, $$
 where, $\nu$ will determine whether the spacetime will represent regular or a singular geometry. The value of $\nu$ for a regular and singular black hole is $4$ and $3$, respectively. Where, $l > 0$ is a new parameter with a dimension of a length. The theory does not specify the value of $l$, although it is reasonable to assume that it is of the Planck length scale order, $l \approx P_{l}$ , or of the order of the black hole mass, $l \approx M$, as these are the only two scales that are already in the model. In this paper, we consider the second scenario with $l$ of the order of M, because it is the only one with observational implications for astrophysical black holes. The line element of the Kerr black hole which can be written as,

\begin{widetext}
\begin{eqnarray}
ds^{2}_{Kerr} = - \bigg(1 - \frac{2 M r }{\Sigma}\bigg) dt^2 - \frac{4 M a  r}{\Sigma} sin^2\theta dt d\phi + \frac{\Sigma}{\Delta} dr^2  + \Sigma d\theta^2 + \bigg(a^2 + r^2 + \frac{2 M r a^2 sin^2\theta}{\Sigma}\bigg)  sin^2\theta  d\phi^2, 
\label{kerrlineele}
\end{eqnarray}
\end{widetext}
where, $ \Delta = r^2 - 2 M r + a^2 $. From Eq.\,(\ref{nuvalueeqn}) the metric tensor components for regular and a singular black hole is written as,
$$ g_{tt} = -\left(1 + \frac{l^{2}}{\Sigma}\right)^{\nu} \bigg(1 - \frac{2 M r }{\Sigma}\bigg), $$
$$ g_{rr} = \left(1 + \frac{l^{2}}{\Sigma}\right)^{\nu}  \frac{\Sigma}{\Delta}, $$
$$ g_{\theta \theta} = \left(1 + \frac{l^{2}}{\Sigma}\right)^{\nu} \Sigma, $$
$$ g_{\phi \phi} = \left(1 + \frac{l^{2}}{\Sigma}\right)^{\nu} \bigg(a^2 + r^2 + \frac{2 M r a^2 sin^2\theta}{\Sigma}\bigg)  sin^2\theta, $$
$$ g_{t\phi} = - \left(1 + \frac{l^{2}}{\Sigma}\right)^{\nu}  \frac{4 M a  r}{\Sigma} sin^2\theta. $$

The coordinate singularity in both spacetime can be defined using $\Delta = 0$,
\begin{equation}
    r_{\pm} =  \left(M \pm \sqrt{M^2 - a^2}\right),\\
    \label{ehrotatingct}
\end{equation}
where $r_{\pm}$ referees to outer and inner horizons. The ergoregion in both spacetimes can be determine using $g_{tt} = 0$,
\begin{equation}
    r_{erg\pm} = \frac{1}{2}\left(2 M + \sqrt{2}\sqrt{2 M^{2} -  a^{2} \left(1 - cos(2\theta)\right)} \right).
    \label{singureguergo}
\end{equation}
One may note that the expressions for outer/inner horizons and the ergoregion are same as in Kerr black hole. We use the Eqs.\,(\ref{omegapm}), (\ref{omegazero}), (\ref{engeffi}) to determine extracted energy efficiency rate in a regular and a singular geometries. The angular velocity for both black hole spacetimes with respect to an asymptotic observer is written as,
\begin{widetext}
\begin{eqnarray}
    \Omega_{\pm} =  \frac{2 a M r}{a^2 \sin ^2(\theta ) \left(a^2 \cos ^2(\theta )+r (2 M+r)\right)+\left(a^2 \cos ^2(\theta )+r^2\right)^2} \pm \sqrt{ J - K},
\end{eqnarray}
where, $$ J =  \frac{2 a M r}{a^2 \sin ^2(\theta ) \left(a^2
   \cos ^2(\theta )+r (2 M+r)\right)+\left(a^2 \cos ^2(\theta )+r^2\right)^2},$$ and
\end{widetext}
   $$ K = \frac{\csc ^2(\theta ) \left(a^2 \cos ^2(\theta )+r (r-2 M)\right)}{2 a M r}. $$

Here in both black hole cases, the ergoregion is same as what we have in the Kerr black hole which one can understand from the mathematical expression defining that region. The ergoregion shows significantly evident changes for the case $a > M$ and $a < M.$ However, we consider only the case where $a < M$ for which the ergoregion exists. The changing of outer/inner horizons and ergoregion from $\theta = \pi/2$ to $\theta = 0$ with different spin parameter are shown in Fig.\,(\ref{regular}). At $\theta = 0$, a spin effect of the object is the same as the Schwarzschild black hole, where the boundary of ergoregion coincides with the horizons. On the other hand, the maximum effect of objects' spin can be perceived at $\theta= \pi/2.$ All of this can be visualized in Fig.\,(\ref{regular}). For extreme spin (where mass and spin are equal) the Cauchy and event horizons coincides which can be seen in Fig.\,(\ref{a = 1}). The radius of the inner horizon is decreasing as objects spin decreasing. Moreover, as opposed to that, the radius of the outer horizon is increasing with decreasing objects' spin. The shape of an ergoregion also changes as the radius of the inner/outer horizons changes with the spin parameter (note that for slow rotation, where the spin parameter is half of the mass). As the size of ergoregion is changing, the efficiency for energy extraction will change.
  
\begin{figure}
    \centering
    \includegraphics[width=8.5cm]{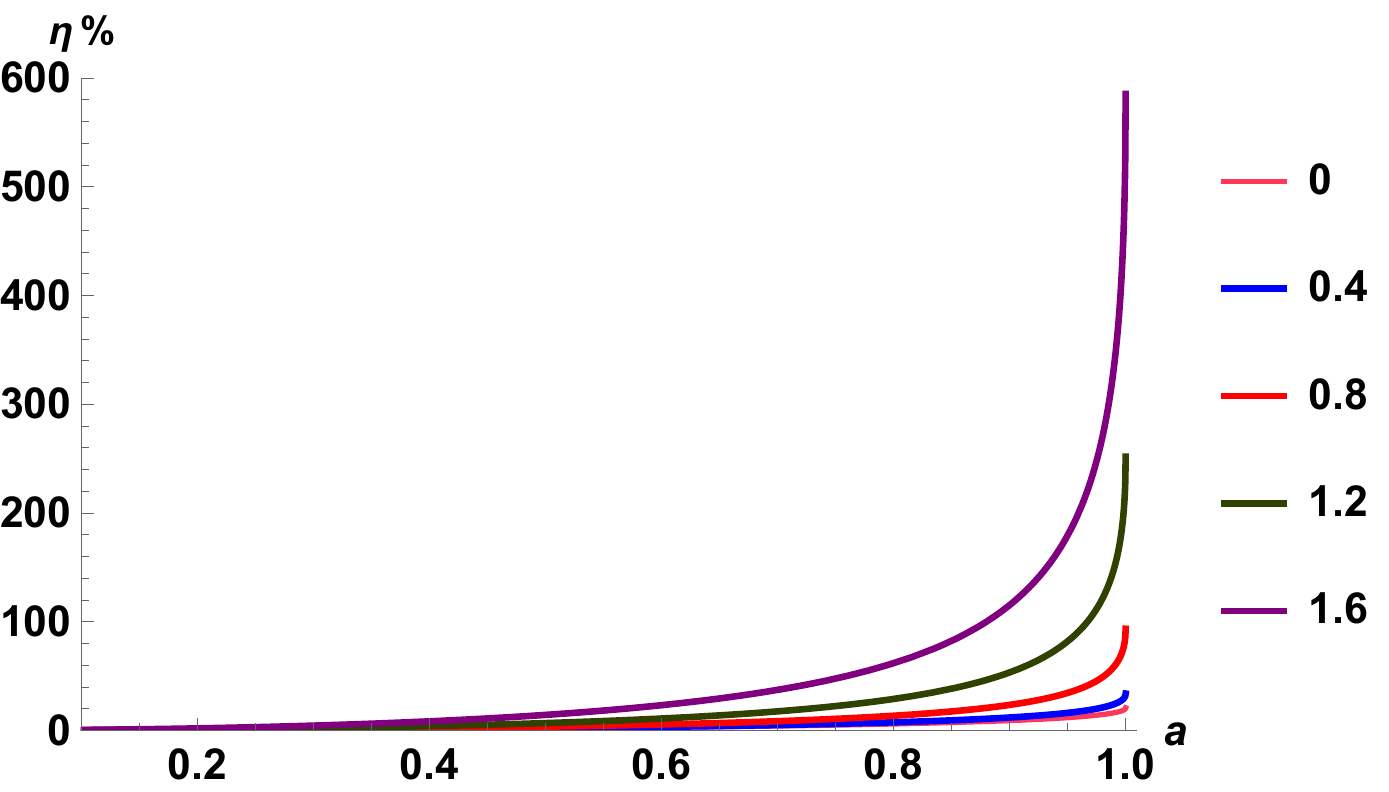}
    \caption{Figure represents the energy extraction efficiency vs spin parameter for a regular black hole. The bar on the right side of page besides figure represents the values of regularisation parameter $l$. where, $l = 0$ is for the Kerr black hole.}
    \label{regulareffiplot}
\end{figure}  
\subsection{A regular black hole}
\label{regularsection}
Let us first look at the regular black hole solution in (\ref{nuvalueeqn}), which can be obtained from the Kerr metric after using the rescaling factor (\ref{rescalingfac}). The line element for a regular black hole is written as,

\begin{widetext}
\begin{eqnarray}
    ds^{2}_{reg} = \left(1 + \frac{l^{2}}{\Sigma}\right)^{4} \left(- \bigg(1 - \frac{2 M r }{\Sigma}\bigg) dt^2 - \frac{4 M a  r}{\Sigma} sin^2\theta\; dt d\phi + \frac{\Sigma}{\Delta} dr^2
    + \Sigma d\theta^2 + \bigg(a^2 + r^2 + \frac{2 M r a^2 sin^2\theta}{\Sigma}\bigg)  sin^2\theta  d\phi^2 \right),
\end{eqnarray}
\end{widetext}

where, the 'reg' refers to the regular black hole and the metric tensor components with $\nu = 4$ are,

\newpage
\begin{widetext}
$$ g_{tt_{(reg)}} = -\left(1 + \frac{l^{2}}{(r^2 +  a^2 cos^{2}\theta)}\right)^{4} \bigg(1 - \frac{2 M r }{(r^2 +  a^2 cos^{2}\theta)}\bigg), $$
$$ g_{rr_{(reg)}} = \left(1 + \frac{l^{2}}{(r^2 +  a^2 cos^{2}\theta)}\right)^{4}  \frac{(r^2 +  a^2 cos^{2}\theta)}{( r^2 - 2 m r + a^2)}, $$
$$ g_{\theta \theta_{(reg)}} = \left(1 + \frac{l^{2}}{(r^2 +  a^2 cos^{2}\theta)}\right)^{4} (r^2 +  a^2 cos^{2}\theta), $$
$$ g_{\phi \phi_{(reg)}} = \left(1 + \frac{l^{2}}{(r^2 +  a^2 cos^{2}\theta)}\right)^{4} \bigg(a^2 + r^2 + \frac{2 M r a^2 sin^2\theta}{(r^2 +  a^2 cos^{2}\theta)}\bigg)  sin^2\theta, $$
$$ g_{t\phi_{(reg)}} = - \left(1 + \frac{l^{2}}{(r^2 +  a^2 cos^{2}\theta)}\right)^{4}  \frac{4 M a  r}{(r^2 +  a^2 cos^{2}\theta)} sin^2\theta. $$
\end{widetext}
Using Kretschmann scalar one may get to know about existence of spacetime singularity. The Kretschmann scalar can be represent with the parameters of Riemann curvature tensor as,
\begin{equation}
    K = R_{a b c d}\, R^{a b c d},
\end{equation}
For regular black hole spacetime the Kretschmann scalar has the form,
\begin{equation}
    K = \frac{1}{\left( \Sigma + l^{2} \right)^{n}} (Polynomial, r, cos\theta, M, a, l),
\end{equation}
where, n represents the integer number. This expression for Kretschmann scalar is everywhere regular for $l \neq 0$. which means that K never diverges. In the case when $ l = 0, $ We revive the well-known Kerr metric with the fact that the Kretschmann scalar diverges at $r \to 0$ with $\theta = \pi/2.$

\begin{table}[]
    \centering  \scalebox{0.80}{%
    \begin{tabular}{||c c c c c c c c c c  ||} 
  
 \hline
 No & Spin Parameter (a) &  l = 0 & l = 0.4 &  l = 0.8 & l = 1.2 & l = 1.6   \\ [0.4ex] 
 \hline
 1 & 0.1 & 0.0627 & 0.0706 & 0.0981 & 0.1583 & 0.4558     \\ 
 \hline
 2 &  0.2 &   0.2544 & 0.2868 & 0.4000  &  0.6480 &  1.8704    \\
 \hline
 3 & 0.3 & 0.5859   &  0.6621 &   0.9295    &  1.5163 & 4.3963       \\ 
 \hline
 4 & 0.4 & 1.0774 & 1.2226  &    1.7331   &   2.8560  &  8.3360       \\
 \hline
 5 & 0.5 & 1.7638  &  2.0133  & 2.8939   &  4.8358  &  14.2433     \\ 
 \hline
 6 & 0.6  &   2.7046  & 3.1130  &  4.5622    &   7.7680  &   23.161      \\
 \hline 
 7 & 0.7 &  4.0084  & 4.6698  &   7.0349    &   12.2844   &   37.2457     \\
 \hline
 8 & 0.8 & 5.9017    & 7.0058    &  10.9976    &  19.887   & 61.8034      \\
 \hline
  9 & 0.9 & 9.0098  &   11.0657    &   18.6299 &  35.51   &  115.227     \\ 
 \hline
 10 & 0.93  &   10.466  & 13.0792  &  22.7707    &   44.3966  &   147.24    \\
 \hline 
 11 & 0.96 &   12.5  &  16.0286  &   29.2575   & 58.7556   &   201.265      \\
 \hline
 12 & 0.99 &   16.1956  & 21.8635  &  43.5115    &    91.6925 &  334.925       \\
 \hline
 13 & 1 &  20.7107  &  30.014  &   66.3072   &  147.02  &   585.65      \\
 \hline
\end{tabular}}
\label{regulareffi}
    \caption{In the given table the efficiency of energy extraction is calculated in a regular black hole spacetime for different $l (0.4, 0.8, 1.2, 1.6)$ and the comparison with the Kerr black hole is given (l = 0). }
\end{table}
The maximum energy extraction efficiency (when splitting happens at the event horizon) can be extracted from regular black hole is explored using the Eq.\,(\ref{engeffi}) with taking $\theta = \pi/2$ ,
\begin{widetext}
\begin{eqnarray}
  \eta_{max(reg)} = \left(\frac{2 a^2 M^2 \left(l^2+r^2\right)^4}{r^9 \left(a^2 (2 M+r)+r^3\right) \sqrt{1-\frac{\left(l^2+r^2\right)^4 (r-2 M)}{r^9}}}+\frac{\sqrt{1-\frac{\left(l^2+r^2\right)^4 (r-2
   M)}{r^9}}+1}{2 \sqrt{1-\frac{\left(l^2+r^2\right)^4 (r-2 M)}{r^9}}}-1\right) 100.
\end{eqnarray}
\end{widetext}

\begin{figure*}
\centering

\subfigure[a = 1.]
{{\includegraphics[width=5cm]{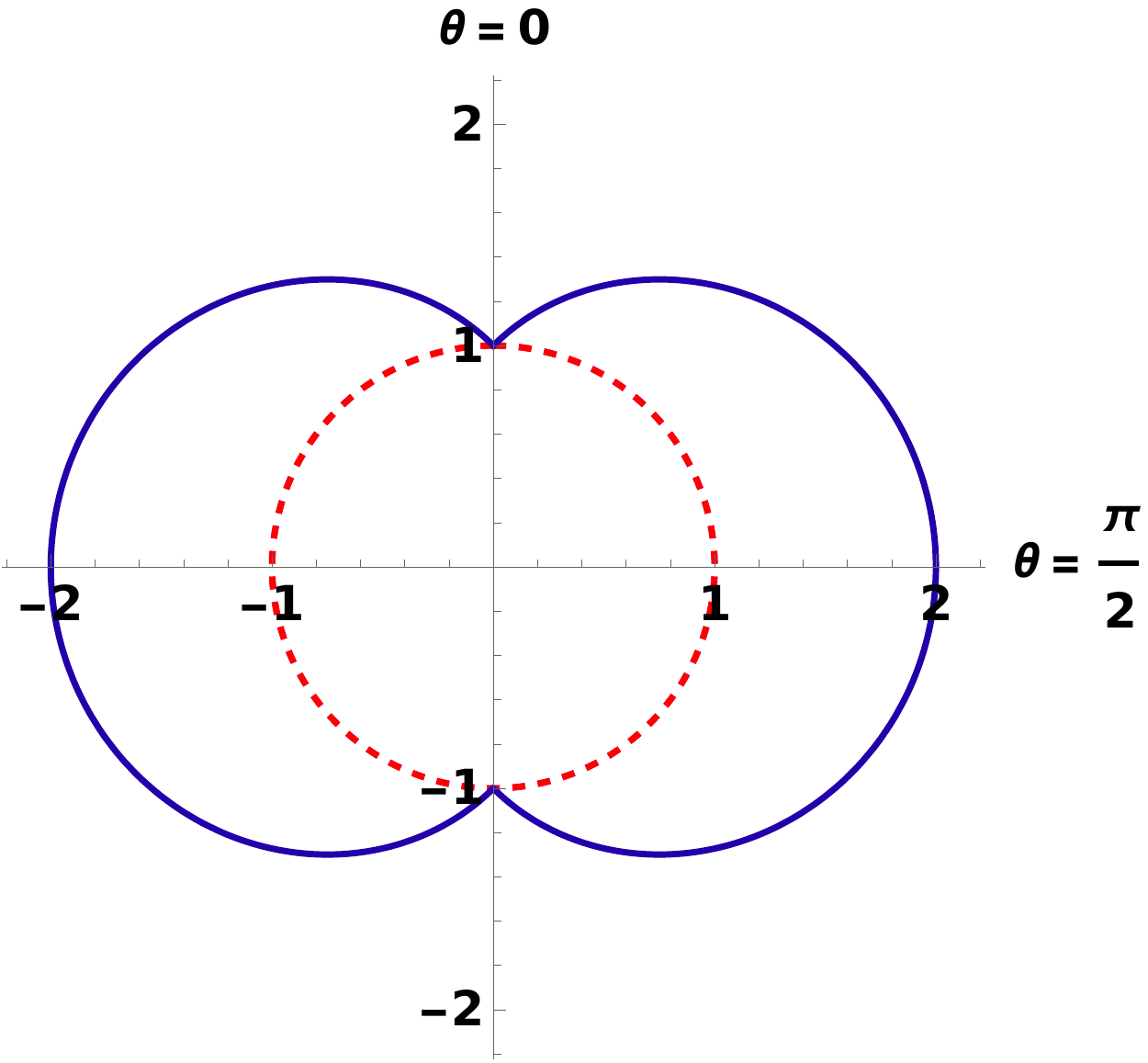}}
\label{a = 1}}
\hspace{0.2cm}
\subfigure[a = 0.99.]
{{\includegraphics[width= 5cm]{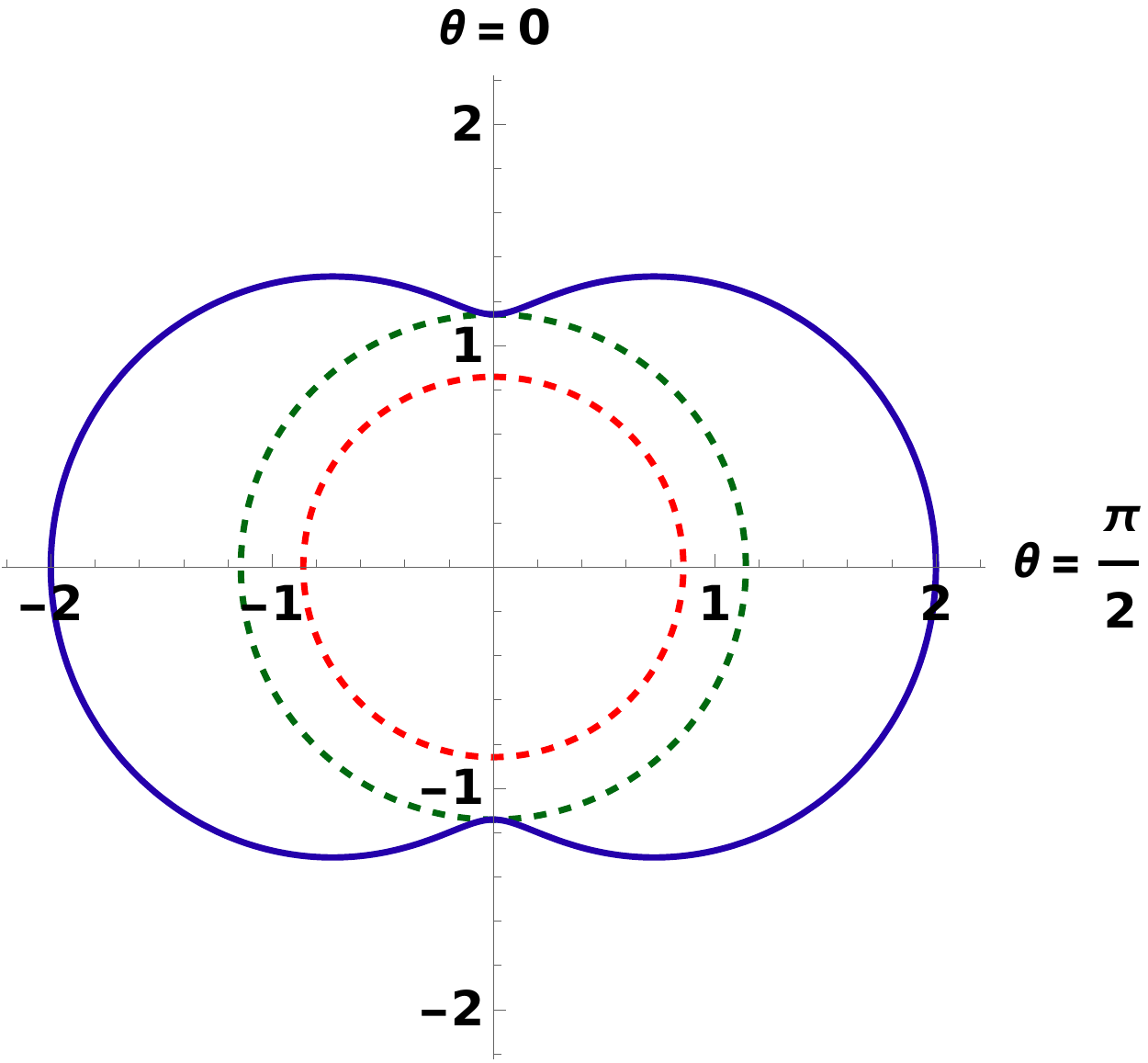} }
\label{a = 0.99}}
\hspace{0.2cm}
\subfigure[a = 0.96.]
{{\includegraphics[width=5cm]{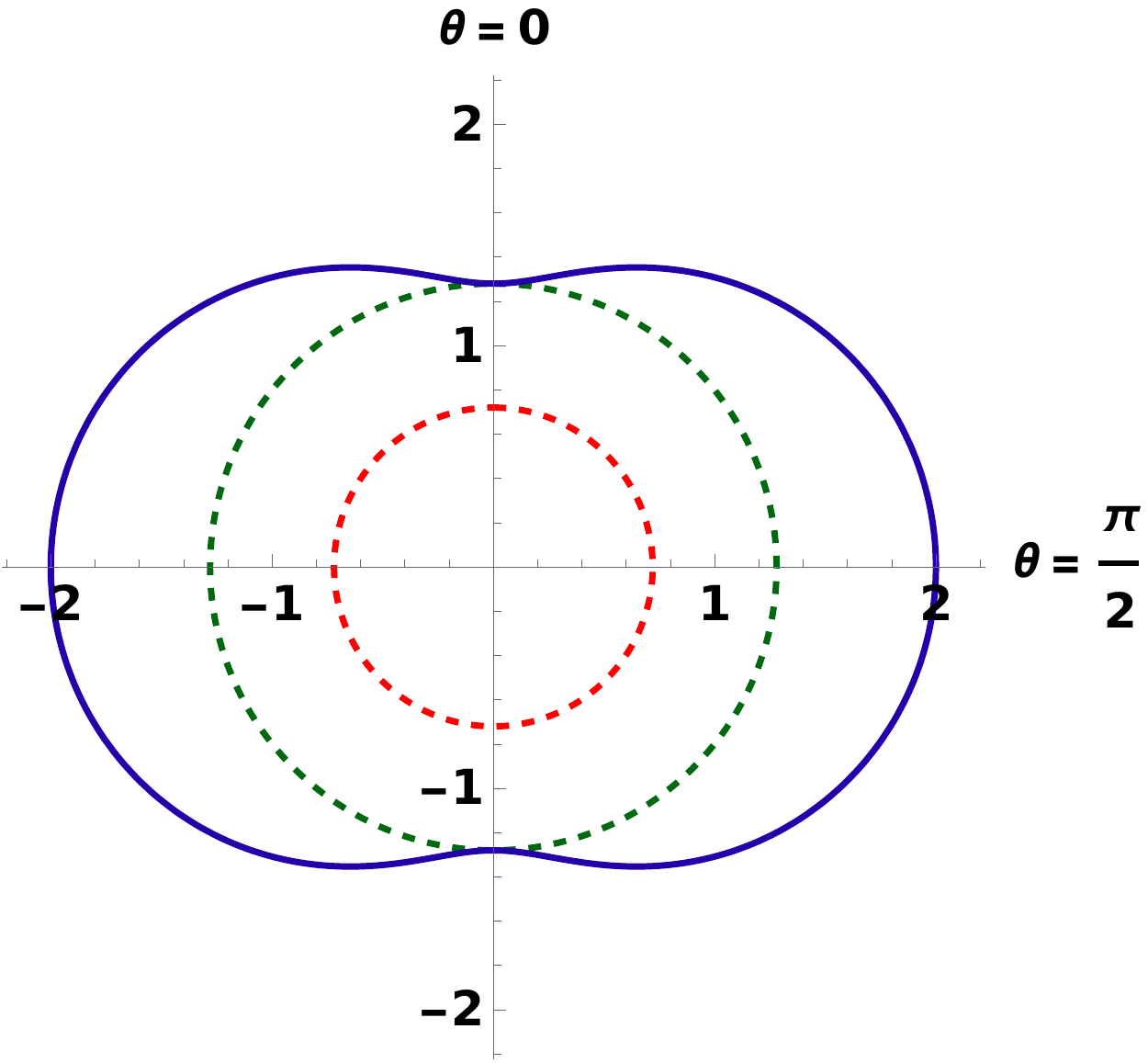}}
\label{a = 0.96}}
\hspace{0.2cm}   
\subfigure[a = 0.93.]
{{\includegraphics[width=5cm]{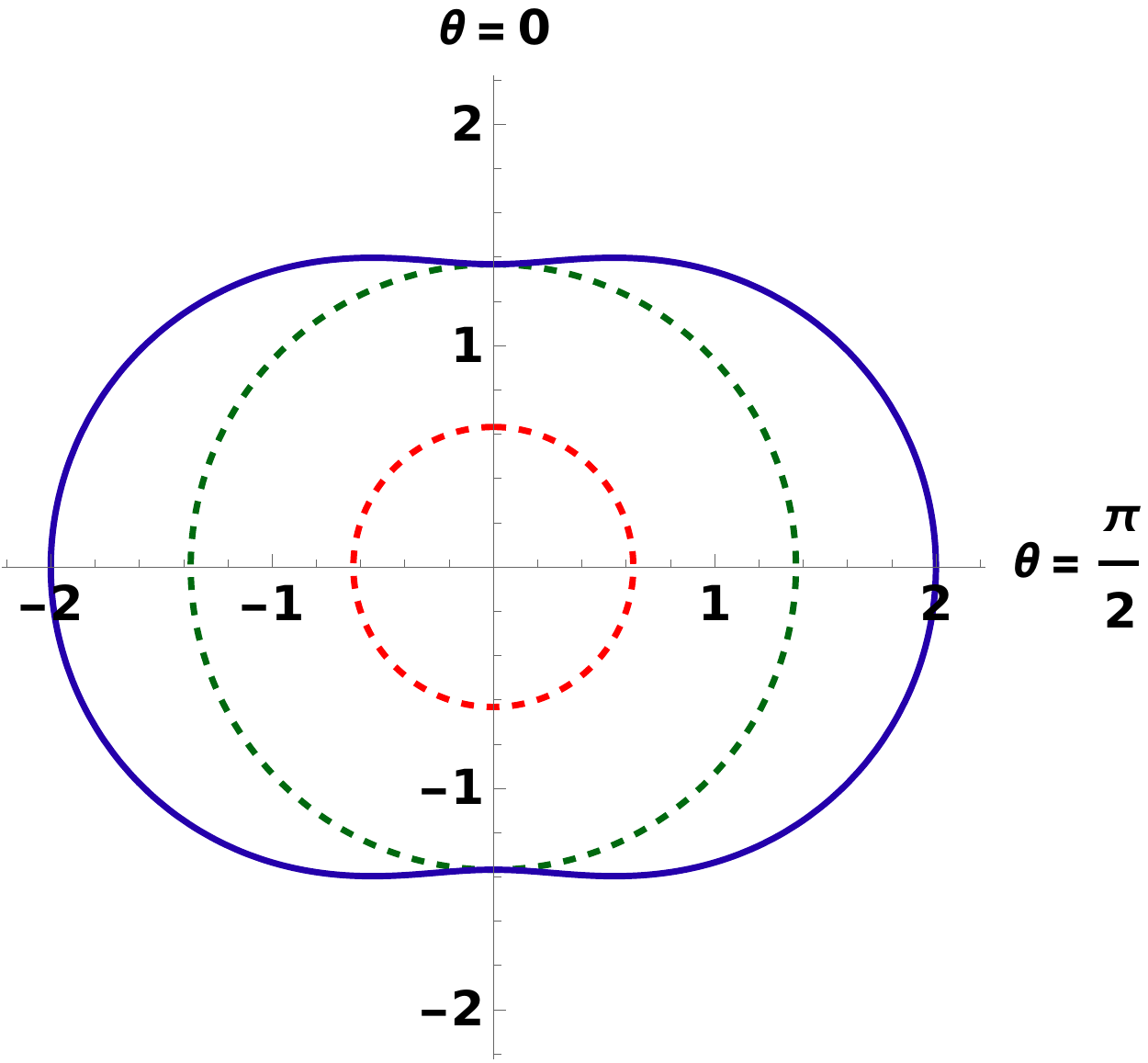}}
\label{a = 0.93}}
\hspace{0.2cm}
\subfigure[a = 0.9.]
{{\includegraphics[width=5cm]{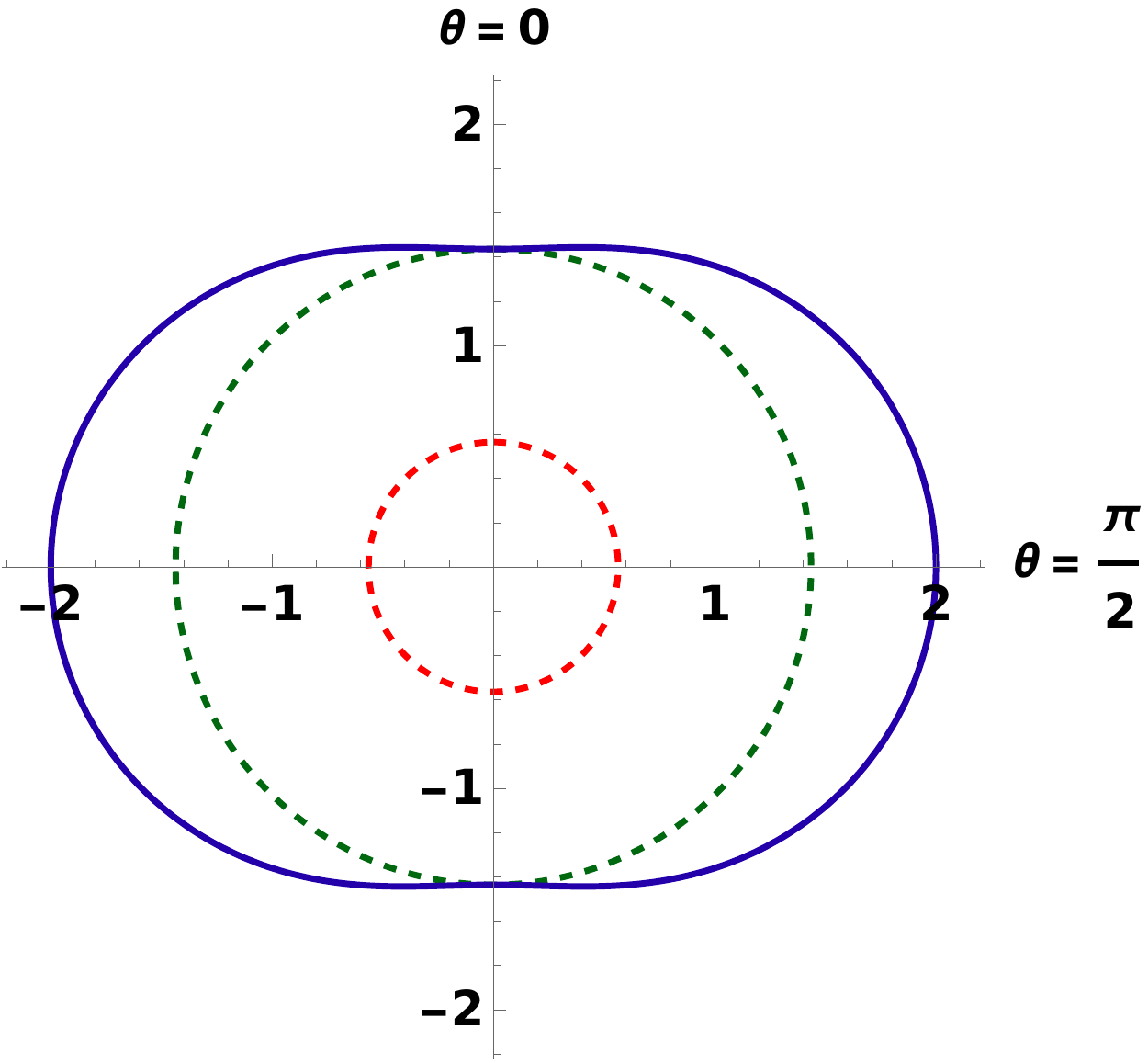}}
 \label{a = 0.9}}
\hspace{0.2cm}
\subfigure[a = 0.8.]
{{\includegraphics[width=5cm]{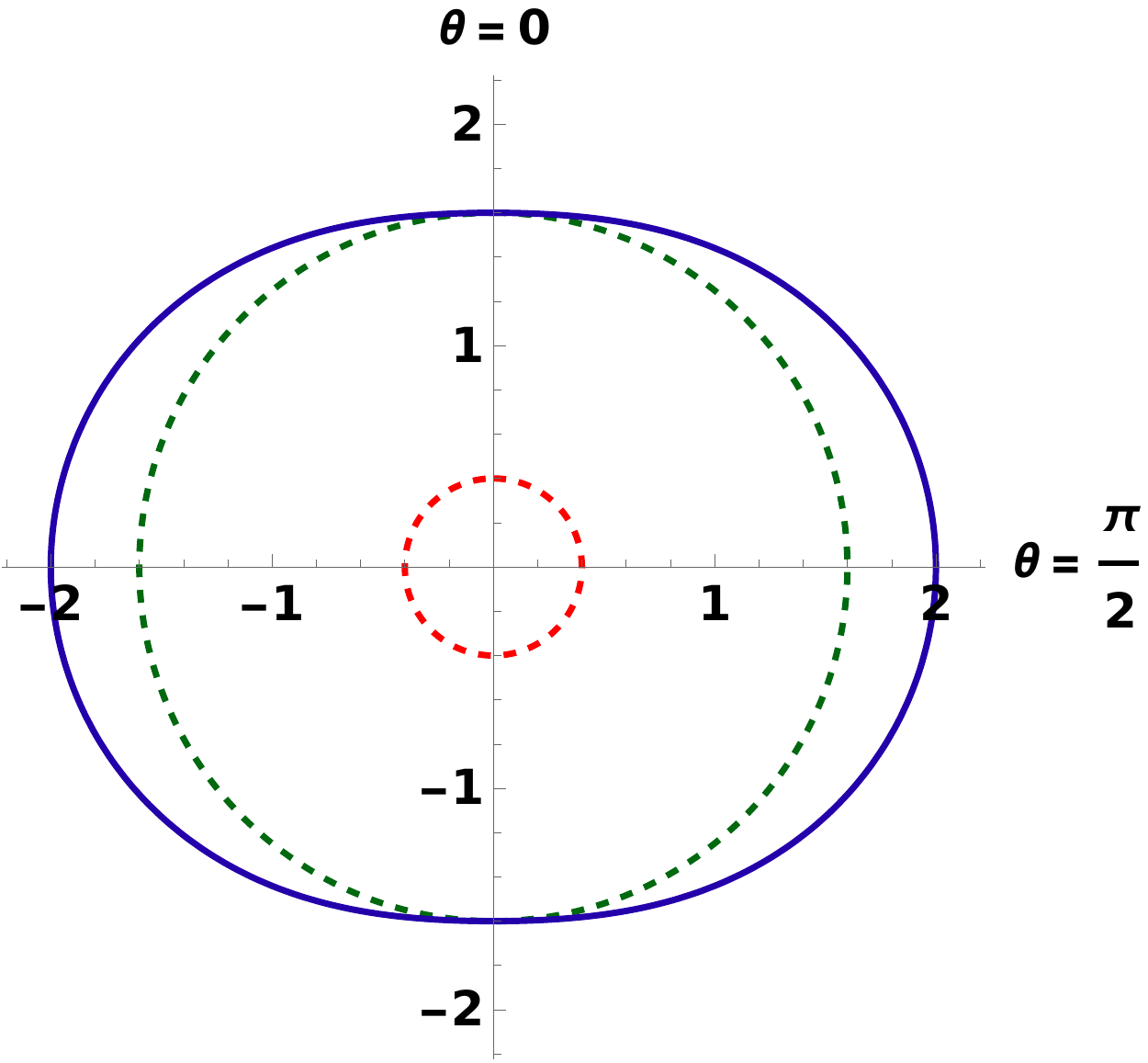}}
\label{a = 0.8}} 
\hspace{0.2cm}
\subfigure[a = 0.7.]
{{\includegraphics[width=5cm]{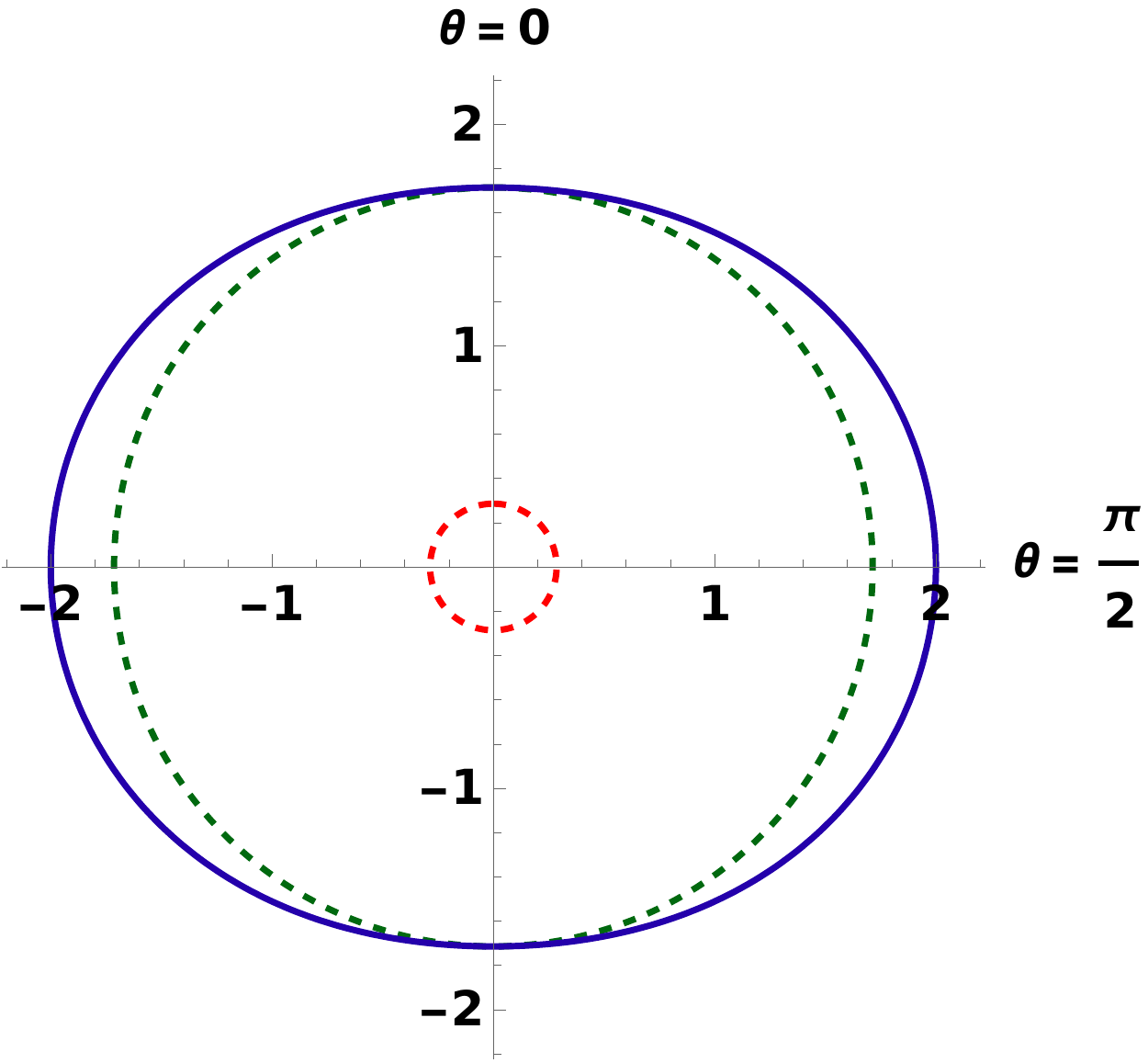}}
\label{a = 0.7}}
\hspace{0.2cm}
\subfigure[a = 0.6.]
{{\includegraphics[width=5cm]{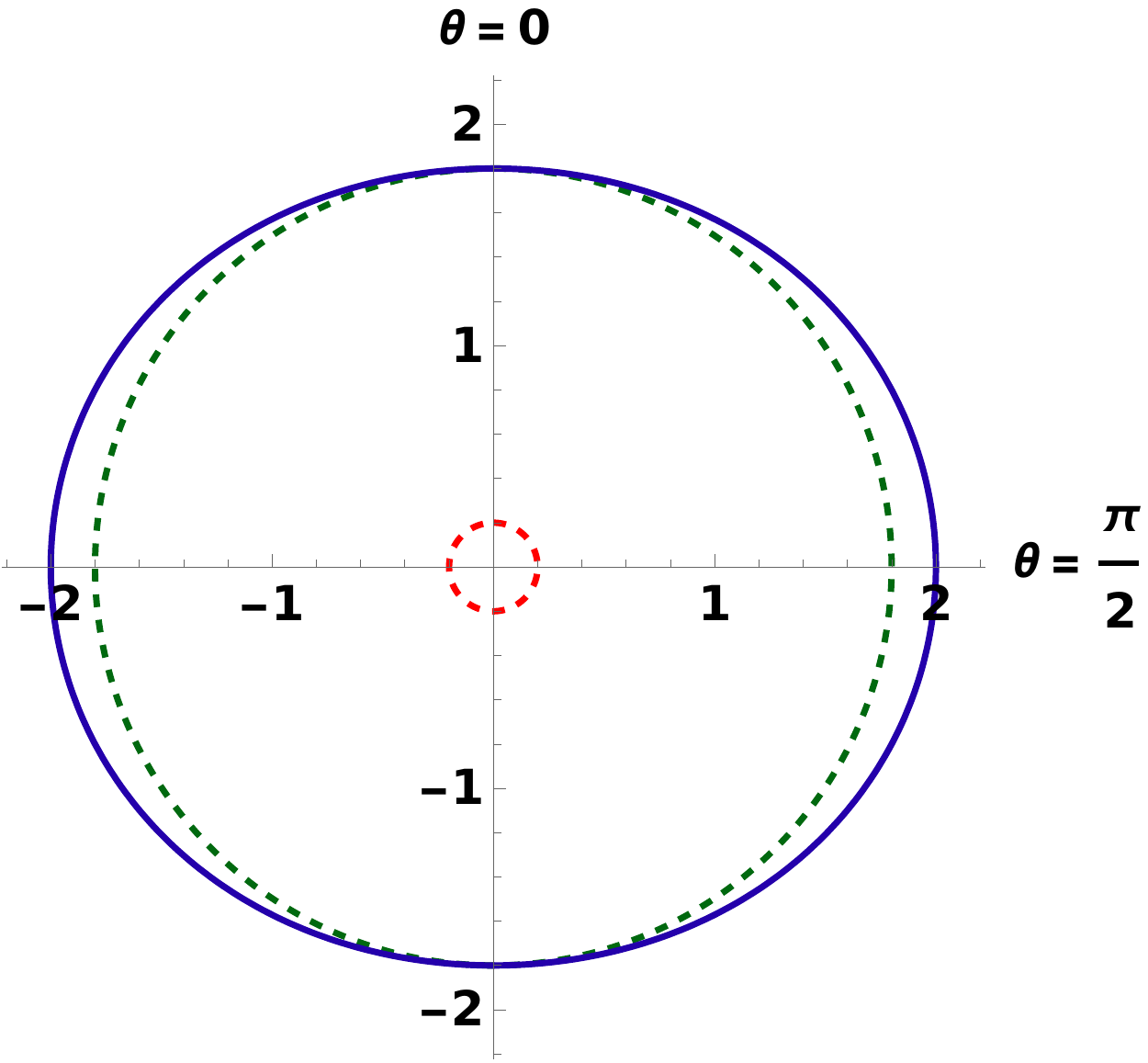}}
\label{a = 0.6}}
\hspace{0.2cm}
\subfigure[a = 0.5.]
{{\includegraphics[width=5cm]{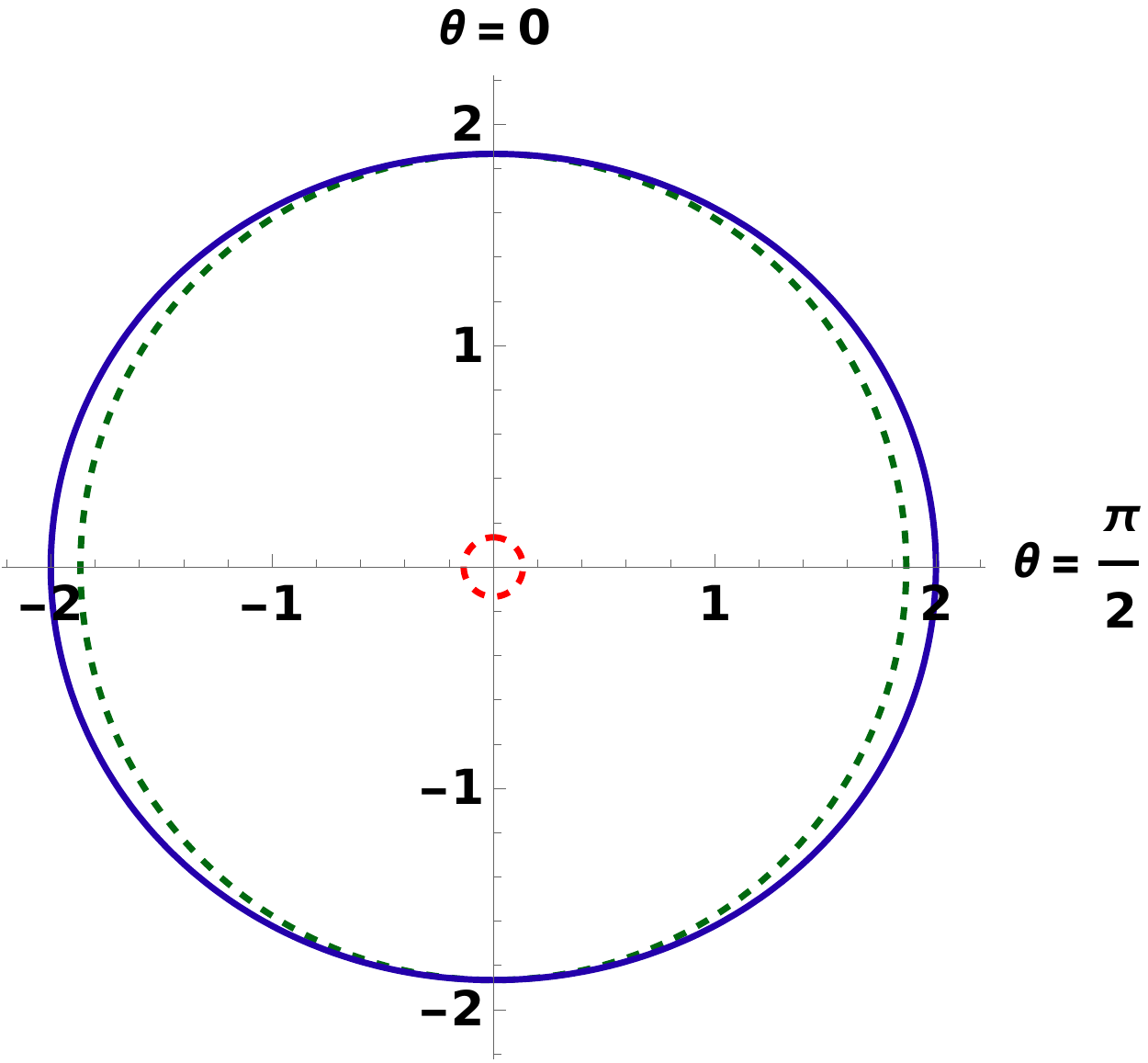}}
\label{a = 0.5}}
\hspace{0.2cm}

\caption{Figures shows the behaviour of ergoregion and event horizon in regular and a singular black hole spacetimes with different parameters. The blue color circle is for outer boundary of ergoregion and the dotted green and red color circle represents the outer and inner horizon respectively. The mass is set equal to one for all the different parameters.  } \label{regular}
\end{figure*}

The table (\ref{regular}) represents the energy extraction efficiency for a regular black hole. It is shown with different spin parameter ($a$) and different values of the regularisation parameter ($l$), where $l = 0$ is for the Kerr black hole. The maximum efficiency of energy extracted in the Kerr black hole at extreme spin is $20.71 \%$, which is the well- known result for a rotating black hole. Whereas for a regular black holes it could be greater than the Kerr black hole. With increasing regularisation parameter ($l$) the energy extraction efficiency is increasing in the regular black hole, as can be seen in (\ref{regular}). The maximum efficiency of energy extraction in the regular black hole at extreme spin is $585.65 \%$ for $l = 1.6$. The variation between extracted energy with $l = 0$ and $l = 1.6$ is comparably minimal at slow rotation (where the spin parameter is half of the mass), as it is substantially larger for the high spin parameter as shown in Fig.\,(\ref{regulareffiplot}). As noted previously, the ergoregion is maximum at the extreme objects' spin, resulting the maximum energy extraction efficiency. The ergoregion reduces with decreasing spin parameter causing the reduction of energy extraction efficiency. 

\subsection{A singular black hole  }
\label{singularsection}
Now considering the singular black hole solution in (\ref{nuvalueeqn}), which can be obtain from the Kerr metric after using $ \nu = 3 $ in the rescaling factor (\ref{rescalingfac}). After that, the line element for a singular black hole can be written as,
\begin{widetext}
\begin{eqnarray}
    ds^{2}_{sing} = \left(1 + \frac{l^{2}}{\Sigma}\right)^{3} \left(- \bigg(1 - \frac{2 M r }{\Sigma}\bigg) dt^2 - \frac{4 M a  r}{\Sigma} sin^2\theta\; dt d\phi + \frac{\Sigma}{\Delta} dr^2  + \Sigma d\theta^2 + \bigg(a^2 + r^2 + \frac{2 M r a^2 sin^2\theta}{\Sigma}\bigg)  sin^2\theta  d\phi^2 \right),
\end{eqnarray}
\end{widetext}
where, the 'sing' refers to the singular black hole and the metric tensor components can be written as,
\newpage
\begin{widetext}
\begin{eqnarray}
&& g_{tt_{(sing)}} = -\left(1 + \frac{l^{2}}{(r^2 +  a^2 cos^{2}\theta)}\right)^{3} \bigg(1 - \frac{2 M r }{(r^2 +  a^2 cos^{2}\theta)}\bigg),\\
\vspace{2cm}
 && g_{rr_{(sing)}} = \left(1 + \frac{l^{2}}{(r^2 +  a^2 cos^{2}\theta)}\right)^{3}  \frac{(r^2 +  a^2 cos^{2}\theta)}{( r^2 - 2 M r + a^2)}, \\
 && g_{\theta \theta_{(sing)}} = \left(1 + \frac{l^{2}}{(r^2 +  a^2 cos^{2}\theta)}\right)^{3} (r^2 +  a^2 cos^{2}\theta), \\
 && g_{\phi \phi_{(sing)}} = \left(1 + \frac{l^{2}}{(r^2 +  a^2 cos^{2}\theta)}\right)^{3} \bigg(a^2 + r^2 + \frac{2 M r a^2 sin^2\theta}{(r^2 +  a^2 cos^{2}\theta)}\bigg)  sin^2\theta, \\
&& g_{t\phi_{(sing)}} = - \left(1 + \frac{l^{2}}{(r^2 +  a^2 cos^{2}\theta)}\right)^{3}  \frac{4 M a  r}{(r^2 +  a^2 cos^{2}\theta)} sin^2\theta. 
 \end{eqnarray}
\end{widetext}
The maximum efficiency can be extracted from a singular black hole can be explore using the Eqn. (\ref{engeffi}) with taking $\theta = \pi/2$ ,
\begin{widetext}
\begin{eqnarray}
  \eta_{max(sing)} = \left(\frac{2 a^2 M^2 \left(l^2+r^2\right)^3}{r^7 \left(a^2 (2 M+r)+r^3\right) \sqrt{1-\frac{\left(l^2+r^2\right)^3 (r-2 M)}{r^7}}}+\frac{\sqrt{1-\frac{\left(l^2+r^2\right)^3 (r-2
   M)}{r^7}}+1}{2 \sqrt{1-\frac{\left(l^2+r^2\right)^3 (r-2 M)}{r^7}}}-1\right)  100.
\end{eqnarray}
\end{widetext}

\begin{table}[]
    \centering   \scalebox{0.83}{%
    \begin{tabular}{||c c c c c c c c c c  ||} 
 \hline
 No & Spin Parameter (a) &  l = 0 & l = 0.4 &  l = 0.8 & l = 1.2 & l = 1.6   \\ [0.4ex] 
 \hline
 1 & 0.1 & 0.0627 & 0.0734 & 0.1139 & 0.2155 &   0.2779     \\ 
 \hline
 2 &  0.2 &   0.2544 & 0.2985 & 0.4650  &  0.8840 &  1.1396      \\
 \hline
 3 & 0.3 & 0.5859   &  0.6897 &   1.0836  & 2.0767 & 2.6758         \\ 
 \hline
 4 & 0.4 & 1.0774 & 1.2752  &  2.0292  &   3.9345 &    5.0652      \\
 \hline
 5 & 0.5 & 1.7638  & 2.104  & 3.4087  &  6.7152 &   8.6324      \\ 
 \hline
 6 & 0.6  &   2.7046  & 3.2620 &  5.4190    &  10.9033 &    13.9803        \\
 \hline 
 7 & 0.7 &  4.0084  & 4.9127  &   8.4550   &  17.4932 &   22.3309       \\
 \hline
 8 & 0.8 & 5.9017    & 7.4153   &  13.4498    &  28.9005  & 36.6025      \\
 \hline
  9 & 0.9 & 9.0098  &   11.8417   &  23.4581 &   53.3091 &  66.4615      \\ 
 \hline
 10 & 0.93  &   10.466  &  14.0743  &  29.0711  &  67.6721  &    83.6705     \\
 \hline 
 11 & 0.96 &   12.5  &   17.3899 &    38.0852   &   91.5133   &  111.747        \\
 \hline
 12 & 0.99 &   16.1956  & 24.1065  & 58.6912   &   148.786  &    177.178      \\
 \hline
 13 & 1 &  20.7107  &  33.8248  &   93.4743   &   251.85  &     289.551     \\
 \hline
\end{tabular} }
  \caption{In this table, the efficiency of energy extraction is shown in a singular black hole spacetime for different $l (0.4, 0.8, 1.2, 1.6)$ and  the comparison with the Kerr black hole is given (l = 0). }
    \label{singulartable}
\end{table}

\begin{figure}
    \centering
    \includegraphics[width=9.3cm]{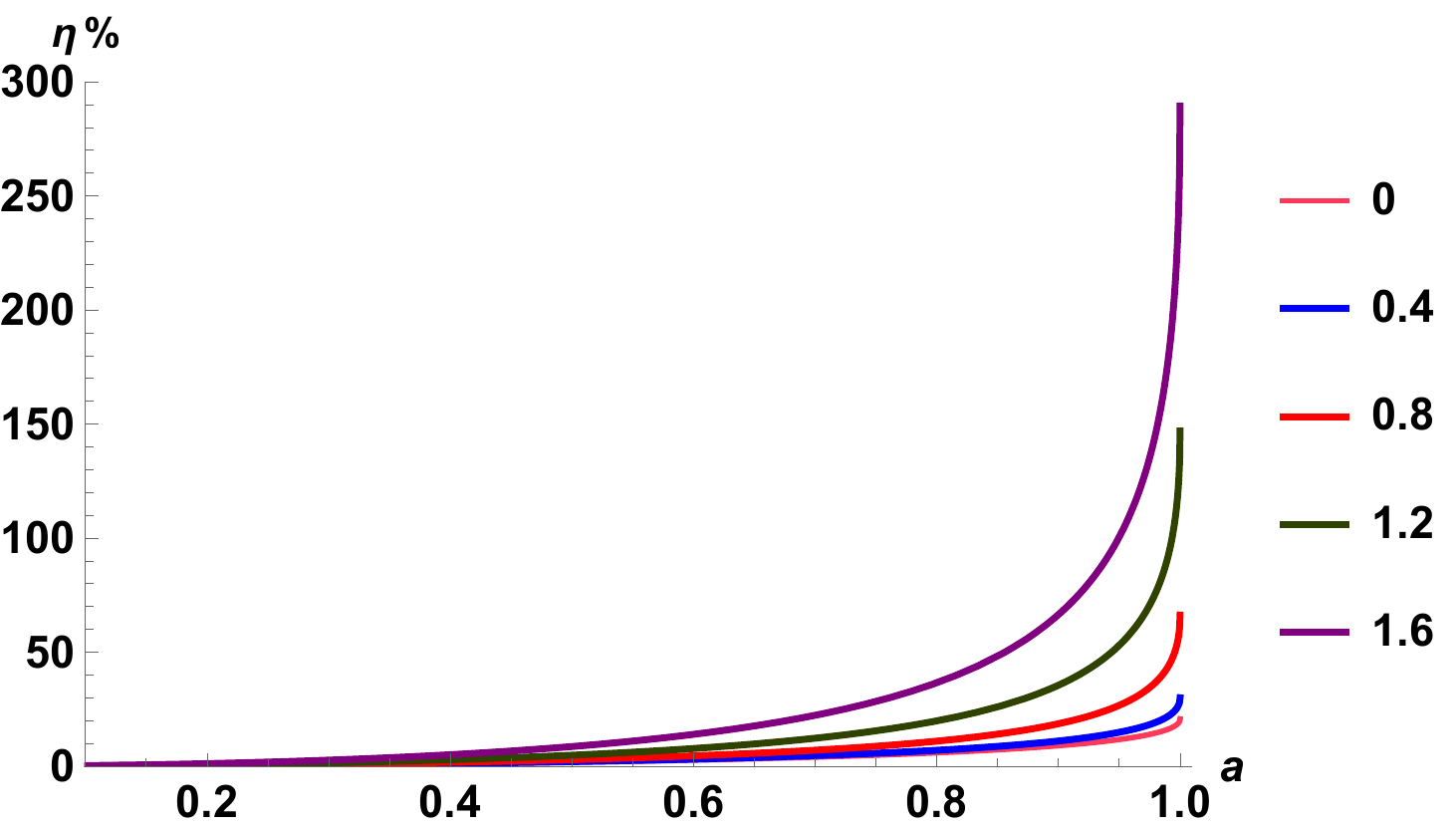}
    \caption{Figure represents the energy extraction efficiency vs spin parameter for a singular black hole. The bar on the right side of page besides figure represents the values of regularisation parameter $l$. where, $l = 0$ is for the Kerr black hole.}
    \label{singulareffiplot}
\end{figure}

The table (\ref{singulartable}) represents the energy extraction efficiency for a singular black hole. It is shown with different spin parameter ($a$) and different values of the regularisation parameter ($l$). where $l = 0$ is for the Kerr black hole. For a singular black hole the efficiency of extracted energy could be greater than the Kerr black hole. With increasing regularisation parameter ($l$), the energy extraction efficiency is increasing in the singular black hole, as can be seen from the table (\ref{singulartable}). For $l = 1.6$, the maximum energy extracted in the singular black hole at extreme spin is $289.55\%$. The variation between extracted energy with $l = 0$ and $l = 1.6$ is comparably minimal at slow rotation (where the spin parameter is half of the mass), as it is substantially larger in the high spin parameter. It is also represented in Fig. (\ref{singulareffiplot}). A regular and singular black hole's energy extraction efficiency is significantly greater than the Kerr black hole case. Note that, a regular black hole, on the other hand, has even more extracted energy than a singular black hole.

\section{Discussion and Conclusions}
\label{conclusion}
 In this paper, we conduct the comparative investigation of the energy extraction using the Penrose process in rotating regular versus singular spacetimes. First, we have discussed the Penrose process. Then we examined the rotating Simpson-Visser spacetime, which has a family of different solutions, and studied how the ergoregion and horizons are changes as the spin and regularisation parameters change. While, in conformal gravity, we investigate the efficiency of energy extraction in a singular and a regular black hole spacetimes and compare it to the Kerr black hole case. 
 The following are the outcomes of this study. 
\begin{itemize}
    \item In the Simpson-Visser spacetime the ergoregion is dependent on the regularisation parameter ($l$). It is evident that the ergoregion and outer/inner horizons show significant changes as the spin parameter and regularisation parameter change. The Penrose process to extract rotational energy from rotating objects is exclusively dependent on the ergoregion and the purpose of this study was to see how the Penrose process might be used to extract the maximum energy from a non-singular compact object such as a wormhole and regular black holes. As the ergoregion and horizons are differ from the Kerr black hole, the efficiency of energy extraction should be different from the Kerr black hole. Unexpectedly, we found that the energy extraction in rotating Simpson-Visser spacetime is same as in the Kerr black hole. That is because, the efficiency of energy extraction ($\eta_{max}$) is independent of the regularisation parameter $l$. 
    
    \item  Using the conformal transformation classically, one can resolve the spacetime singularity problem that arises in Einstein's general theory of relativity. The singular and a regular black holes considered here are solution of CEFE derived in \cite{Bambi:2016wdn}. Depending on the parameter $\nu$ one gets the spacetime solution with and without singularity. The expressions of ergoregions in a singular and a regular black hole spacetimes are independent of the regularisation parameter ($l$). Thus, the ergoregions for a regular and a singular black holes are similar to that of the Kerr black hole. As explained earlier, the ergoregion and horizons show significantly evident changes for the case $a > M$ and $a < M$. However, we consider only the case in which the $a < M$ for that the horizons are exist. 
    
   \item  It is evident from these investigation that the efficiency of energy extraction will vary as the size of the ergoregion changes. Interestingly, even though the ergoregions in a regular and a singular black holes are similar as in the Kerr black hole, the efficiency for energy extraction is significantly larger in regular and a singular black holes. In a CEFE solutions, the efficiency of energy extraction is large enough in a regular black hole rather than in a singular black hole case. However, one may notice from Figs.\,(\ref{regulareffiplot}) and (\ref{singulareffiplot}) that in all compact objects, the energy extraction is nearly the same for spin parameter up to 0.5. The maximum difference for energy extraction efficiency occurs at extreme spin parameter (a = M).   
   
   \item In this work, the phenomenology of energy extraction for a neutral test particle is explained for singular and regular black holes. One may study the efficiency of energy extraction in the presence of a magnetic field or for charge test particle.  
\end{itemize}

\acknowledgments
VP would like to acknowledge the support of the SHODH fellowship (ScHeme Of Developing High quality research - MYSY). Authors would like to express their gratitude towards Divyesh Solanki, Siddharth Madan, Saurabh And Ashwathi Nair for all valuable suggestions and astute insights.


\end{document}